\documentclass[%
 reprint,
 amsmath,amssymb,
 superscriptaddress,
 aps
]{revtex4-2}

\usepackage{graphicx} 
\usepackage{dcolumn} 
\usepackage{bm} 
\usepackage{enumerate}
\usepackage{mathtools}
\usepackage{physics}
\usepackage{booktabs}
\usepackage{soul}
\usepackage{enumitem}

\makeatletter
\let\MYcaption\@makecaption
\makeatother
\usepackage{subcaption}
\makeatletter
\let\@makecaption\MYcaption
\makeatother
\usepackage{hyperref}
\usepackage{xcolor}

\begin{document}



\title{Parameter estimation of protoneutron stars from gravitational wave signals using the Hilbert-Huang transform}

\author{Seiya Sasaoka}
\affiliation{Department of Physics, Tokyo Institute of Technology, 2-12-1 Ookayama, Meguro-ku, Tokyo 152-8551, Japan}%

\author{Yusuke Sakai}
\affiliation{Department of Design and Data Science and Research Center for Space Science, Advanced Research Laboratories, \\ Tokyo City University, 3-3-1 Ushikubo-Nishi, Tsuzuki-ku, Yokohama, Kanagawa 224-8551, Japan}%

\author{Diego Dominguez}
\affiliation{Department of Physics, School of Science, Institute of Science Tokyo, 2-12-1 Ookayama, Meguro-ku, Tokyo 152-8551, Japan}%

\author{Kentaro Somiya}
\affiliation{Department of Physics, School of Science, Institute of Science Tokyo, 2-12-1 Ookayama, Meguro-ku, Tokyo 152-8551, Japan}%

\author{Kazuki Sakai}
\affiliation{Department of Electronic Control Engineering, National Institute of Technology, Nagaoka College, 888 Nishikatakai, Nagaoka, Niigata 940-8532, Japan}

\author{Ken-ichi Oohara}
\affiliation{Graduate School of Science and Technology, Niigata University, 8050 Ikarashi-2-no-cho, Nishi-ku, Niigata 950-2181, Japan}
\affiliation{Niigata Study Center, The Open University of Japan, 754 Ichibancho, Asahimachi-dori, Chuo-ku, Niigata 951-8122, Japan}

\author{Marco Meyer-Conde}
\affiliation{Department of Design and Data Science and Research Center for Space Science, Advanced Research Laboratories, \\ Tokyo City University, 3-3-1 Ushikubo-Nishi, Tsuzuki-ku, Yokohama, Kanagawa 224-8551, Japan}%
\affiliation{University of Illinois at Urbana-Champaign, Department of Physics, Urbana, Illinois 61801-3080, USA}

\author{Hirotaka Takahashi}
\affiliation{Department of Design and Data Science and Research Center for Space Science, Advanced Research Laboratories, \\ Tokyo City University, 3-3-1 Ushikubo-Nishi, Tsuzuki-ku, Yokohama, Kanagawa 224-8551, Japan}%
\affiliation{Institute for Cosmic Ray Research (ICRR), The University of Tokyo, 5-1-5 Kashiwa-no-Ha, Kashiwa City, Chiba 277-8582, Japan}
\affiliation{Earthquake Research Institute, The University of Tokyo, 1-1-1 Yayoi, Bunkyo-ku, Tokyo 113-0032, Japan}

\date{\today}

\begin{abstract}
Core-collapse supernovae (CCSNe) are potential multimessenger events detectable by current and future gravitational wave (GW) detectors.
The GW signals emitted during these events are expected to provide insights into the explosion mechanism and the internal structures of neutron stars.
In recent years, several studies have empirically derived the relationship between the frequencies of the GW signals originating from the oscillations of protoneutron stars (PNSs) and the physical parameters of these stars.
This study applies the Hilbert-Huang transform (HHT) [Proc. R. Soc. A 454, 903 (1998)] to extract the frequencies of these modes to infer the physical properties of the PNSs.
The results exhibit comparable accuracy to a short-time Fourier transform-based estimation, highlighting the potential of this approach as a complementary method for extracting physical information from GW signals of CCSNe.
\end{abstract}

\keywords{}

\maketitle

\section{Introduction\label{sec:intro}}
Advanced gravitational-wave (GW) detectors now regularly observe signals from compact binary coalescences, with over 90 detections recorded to date~\cite{ref:Abbott2019, ref:Abbott2020, ref:Abbott2024, ref:Abbott2023}.
The anticipated next milestone for these detectors is the first observation of the GW signal from a core-collapse supernova (CCSN).
A CCSN is an explosion of a massive star, which leads to the formation of either a neutron star or a stellar-mass black hole.
This explosion usually emits electromagnetic waves, neutrinos, and GWs.
Although the exact explosion mechanism remains unclear, it is widely believed that most stars undergo neutrino-driven explosions~\cite{ref:Janka2012, ref:Mezzacappa2024}.
Current GW detectors, like Advanced LIGO~\cite{ref:Aasi2015}, Advanced Virgo~\cite{ref:Acernese2015}, and KAGRA~\cite{ref:Akutsu2019}, are predicted to detect neutrino-driven CCSN explosions occurring within our Galaxy, reaching a distance of $\sim$ 10 kpc~\cite{ref:Marek2021}.
Coherent network analysis of GW signal~\cite{ref:hayama2015coherent} also highlighted several important hydrodynamic features of CCSN.
Next-generation ground-based GW detectors, such as the Cosmic Explorer~\cite{ref:Abbott2017_ce, ref:Reitze2019} and the Einstein Telescope (ET)~\cite{ref:Punturo2010}, are anticipated to enable the detection of these events from other galaxies, such as the Large Magellanic Cloud.
The kHz-band detector NEMO~\cite{ref:Ackley2020, ref:szczepanczyk2022gravitational} has also been proposed in Australia for studying nuclear physics by observing high-frequency GWs. 
The detection of GW signals and their neutrino counterparts from supernovae yields valuable insights into the inner core of stars, offering crucial information about the explosion mechanism.

Modeling the collapse of the stellar core, the bounce, and the subsequent post-bounce evolution is extraordinarily complicated and computationally expensive.
However, remarkable progress has been made in multidimensional numerical simulations of CCSN explosions over the last decade.
The characteristics of the GW waveform depend on the properties of its progenitor, such as mass, angular velocity, and equation of state, as well as on how the simulations approximate general relativistic effects and treat neutrino transport.
Recent developments in these computations have revealed common characteristics of the CCSN signals.
While the current standard burst GW search pipeline, the coherent WaveBurst (cWB)~\cite{ref:Klimenko2008, ref:Klimenko2016, ref:Drago2021}, does not rely on specific waveform morphologies, studies have begun to utilize known features of GW waveforms in detection algorithms.
In particular, the application of machine learning in detection algorithms~\cite{ref:Astone2018, ref:Iess2020, ref:Chan2020, ref:Lopez2021, ref:Iess2023, ref:Sasaoka2023} has been studied extensively.
Machine learning has also been employed to classify equations of state or explosion mechanisms using simulated gravitational waveforms~\cite{ref:Edwards2021, ref:Perez2022, ref:Mitra2024, ref:Powell2024}.

The GW spectrogram of CCSN explosions highlights a noticeable time evolution of the frequency, increasing monotonically over time.
This so-called gravity-mode, also known as $g$-mode, is believed to be associated with the oscillation of PNSs~\cite{ref:Andersson1998, ref:Sotani2016}.
Characterizing these modes is essential to enable CCSN asteroseismology using GWs (e.g.~\cite{ref:Sotani2019}).
In recent years, numerical simulations have investigated the relationship between the frequency of these modes and the mass and radius of PNSs.
In their studies~\cite{ref:Torres2019, ref:Torres2021}, \citeauthor{ref:Torres2019}, calculated the relations between each $f$/$g$/$p$-mode and the physical parameters of the PNS.
These relations are considered \textit{universal}, as they are independent of the equation of state or the progenitor mass.
\citeauthor{ref:Sotani2021}~\cite{ref:Sotani2021} suggest that once we observe a GW signal from a CCSN explosion, we can infer the physical parameters of the PNS by extracting the frequencies of these modes and solving the universal relations.
In pursuing this goal, \citeauthor{ref:Bizouard2021}~\cite{ref:Bizouard2021} developed a method to track the $g$-mode in a spectrogram to estimate the PNS properties with current or next-generation GW detectors.
\citeauthor{ref:Powell2022}~\cite{ref:Powell2022} proposed a reconstruction approach in the spectrogram domain with Bayesian inference and an asymmetric chirplet model to extract the PNS properties.
\citeauthor{ref:Bruel2023}~\cite{ref:Bruel2023} extended the study in Ref.~\cite{ref:Bizouard2021} by combining the data of a multi-detector network and tracking the $g$-mode with a LASSO regression algorithm.
Moreover, \citeauthor{ref:Lagos2023}~\cite{ref:Lagos2023} used a neural network regression model to estimate the slope of the high-frequency mode in a CCSN.

Accurate extraction of GW frequencies is essential for inferring the physical properties of GW sources. Unlike the short-time Fourier transform (STFT) and wavelet transforms, time-frequency maps produced by the Hilbert-Huang transform (HHT)~\cite{ref:Huang1998} are not constrained by the tradeoff between temporal duration and bandwidth—this absence of limitation results in high resolutions along both the time and frequency axes. 

In this context, the HHT has been explored within the GW research domain, as documented by \citeauthor{ref:Camp2007}~\cite{ref:Camp2007}, who recorded the first application of the HHT in GW analysis.
Numerous studies have highlighted the HHT's efficacy in identifying the frequencies of GW signals in addition to its use in detection algorithms.
This includes postmerger studies of binary neutron star coalescence~\cite{ref:Kaneyama2016, ref:Yoda2023}, binary black hole coalescence~\cite{ref:Sakai2017}, and low-frequency modes induced by neutrino-driven convection and standing accretion shock instability (SASI) in a CCSN~\cite{ref:Takeda2021}.

In this study, we propose an HHT-based approach for estimating $M_{\mathrm{PNS}}/R_{\mathrm{PNS}}^2$ from the $g$-mode GW signal, where $M_{\mathrm{PNS}}$ and $R_{\mathrm{PNS}}$ denote the mass and radius of the PNS, respectively.
Our method integrates the cWB technique~\cite{ref:Klimenko2008, ref:Klimenko2016, ref:Drago2021} and the extraction of the $g$-mode frequencies using the HHT to solve the universal relations from Refs.~\cite{ref:Torres2019, ref:Torres2021}.
We first apply the method to the GW signal-only case, then test it on signals in the ET's simulated detector noise, and perform a comparative evaluation of the estimation accuracy with an approach based on STFT.

This paper is organized as follows.
Section~\ref{sec:data} outlines the data used to evaluate our methodology.
Section~\ref{sec:method} details our approach for parameter estimation of PNSs.
The results are presented in Sec.~\ref{sec:result}.
Finally, we conclude our study in Sec.~\ref{sec:concl}.

\section{Simulation data}\label{sec:data}
Our analysis uses two simulated GW signals generated from different models to validate our parameter estimation method under two distinct conditions. The following will describe the two GW signals, the ET detector, and the preprocessing method.
\subsection{Signal waveform}\label{subsec:waveform}
This work uses the GW signals from the he3.5~\cite{ref:Powell2019} and y20~\cite{ref:Powell2020} models.
Both waveforms derive from three-dimensional numerical simulations conducted with the general-relativistic neutrino hydrodynamics code CoCoNuT-FMT~\cite{ref:Muller2010}.
\begin{itemize}[leftmargin=*]
    \item The he3.5 progenitor model is an ultrastripped star evolved from a helium star with an initial mass of $3.5~M_\odot$. The simulation is stopped at 0.7~s after the supernova core bounce. The GW amplitude reaches its highest at a peak frequency of around 900 Hz.
    \item The y20 progenitor model is a nonrotating, solar metallicity helium star with a mass of $20~M_\odot$. It has a significantly higher mass than the first model, the simulation ends at 1.2 s after core bounce, and the GW amplitude is the strongest at $\sim 600$ Hz.
\end{itemize}

\begin{figure}[h]
    \centering
    \includegraphics[width=\columnwidth]{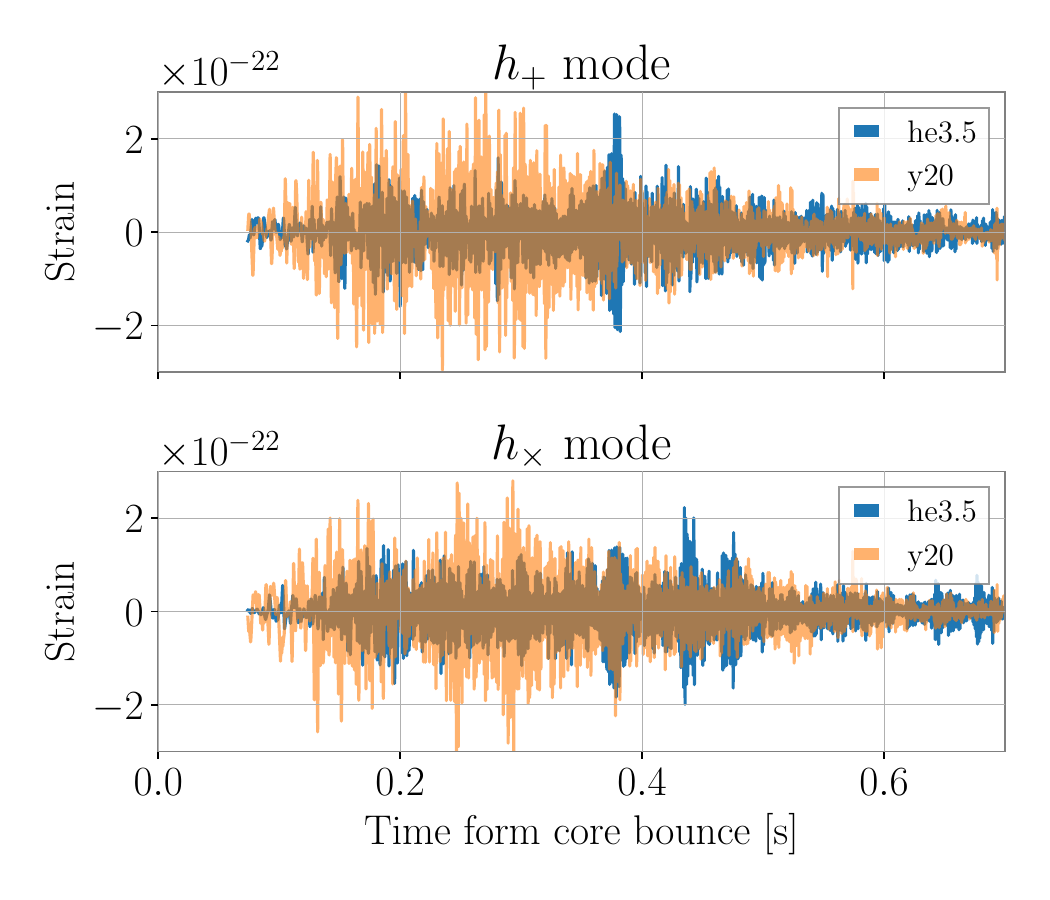}
    \caption{Plus-mode and cross-mode GW signals of he3.5 and y20 models at a source distance of 10 kpc.}
    \label{fig:hphc}
\end{figure}

In this simulation, the GW signal comprises two polarization states: $h_+(t)$ (plus) and $h_\times(t)$ (cross) modes. These modes are shown for each model in Fig.~\ref{fig:hphc}, where the signals are observed from the equatorial plane at a source distance of 10 kpc. Each model exhibits specific features in their signal originating from PNS oscillations.

The strain signal $h(t)$ of a GW signal is given, taking into account the antenna pattern functions $F_{+}$ and $F_\times$:
\begin{align}
    h(t) &= F_{+}(\alpha, \delta, \psi, t)h_{+}(t) + F_{\times}(\alpha, \delta, \psi, t)h_{\times}(t),
\end{align}
where $\alpha$ is the right ascension, $\delta$ is the declination, and $\psi$ is the polarization angle.

\subsection{Einstein Telescope}\label{subsec:ET}
The Einstein Telescope (ET)~\cite{ref:Punturo2010} is a third-generation ground-based GW detector proposed to be constructed underground in Europe.
Its optical layout has yet to be finalized, and various designs have been proposed~\cite{ref:Branchesi2023}.
The reference configuration is based on an equilateral triangle of three detectors, each consisting of two interferometers with 10-km long arms.
One of the two interferometers is specialized for low-frequency bands with low laser power and cryogenic mirrors, while the other is specialized for high-frequency bands with high laser power and mirrors at room temperature.

\begin{figure}[t]
    \centering
    \includegraphics
    [width=0.85\columnwidth]{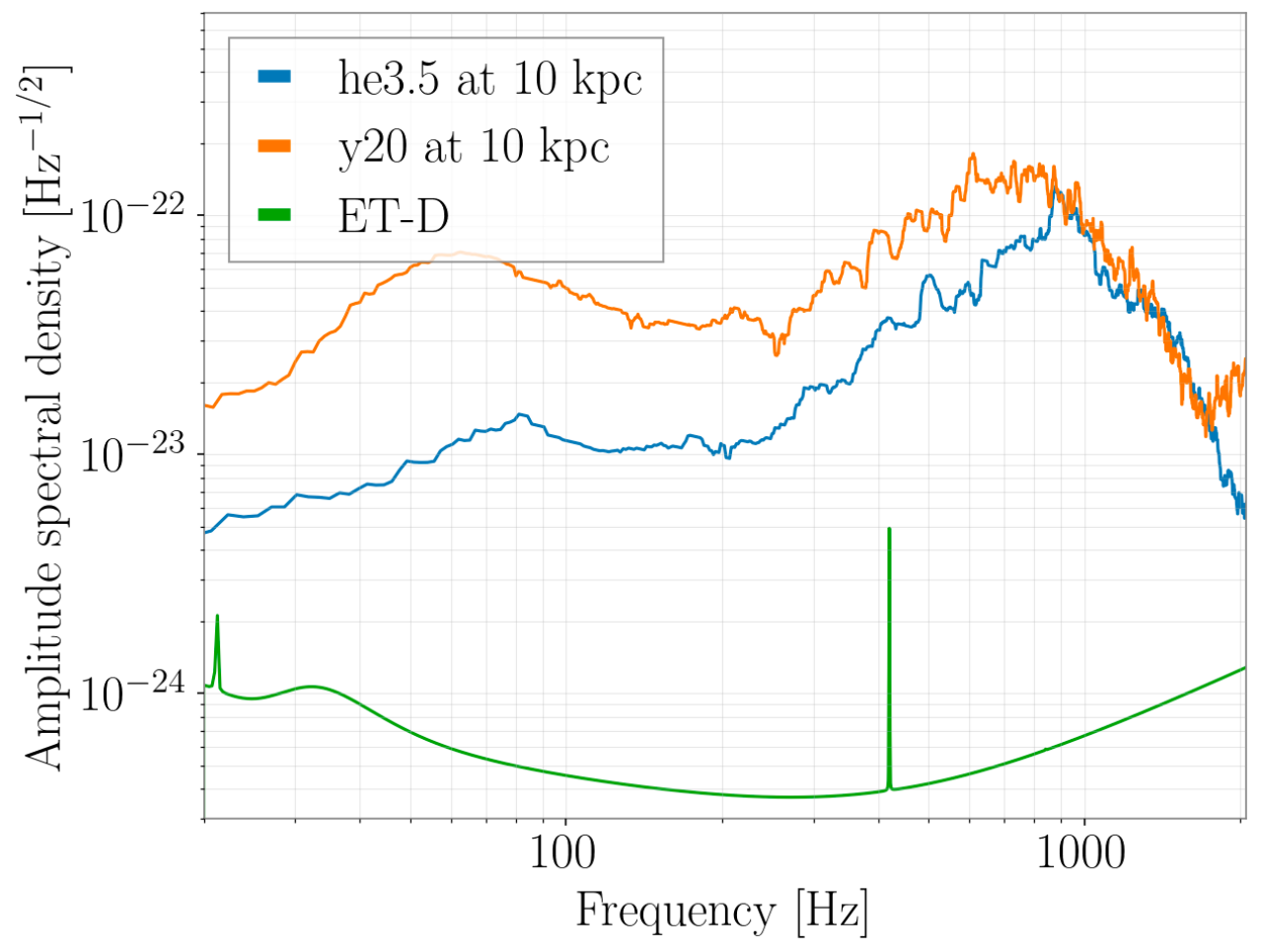}
    \caption{Amplitude spectral density of he3.5 and y20 generated waveforms, along with the ET-D sensitivity curve.}
    \label{fig:asd}
\end{figure}

Figure~\ref{fig:asd} shows the ET-D sensitivity curve~\cite{ref:ET-D} along with the frequency spectrum of the characteristic strain of the he3.5 and y20 waveforms for an observer in the equatorial plane from a distance of 10~kpc.

The characteristic strain $h_c(t)$, is a helpful metric for comparing detector sensitivities with GW signals' detectability~\cite{ref:Marek2021}. Therefore, we use the ET-D sensitivity curve~\cite{ref:Hild2011}, current data analysis standard, to simulate the ET detector noise. Unlike the strain signal $h(t)$ of a GW signal, the characteristic strain provides a way to visualize the potential signal strength relative to the detector's sensitivity ~\cite{ref:moore2014gravitational}. Since the original ET-D sensitivity is computed for an L-shaped interferometer, we normalized by $\sin(60^\circ)$ to make it for a triangular interferometer.

\subsection{Data preprocessing}\label{subsec:preprocess}
For the remainder of this analysis, the observer direction will be fixed in the equatorial plane at all times, and the sky location will be at the ET detector's optimal orientation. We sample GW signals and detector noises at a sampling rate of 4096~Hz and apply a 100-1200~Hz bandpass filter. 
Additionally, we apply a notch filter around 420 Hz to eliminate noise, which appears as a sensitivity peak in Fig.~\ref{fig:asd}. The time origin is set to the moment when the GW amplitude is maximum. Moreover, parts of these computations were carried out using the PyCBC software~\cite{ref:Pycbc}.

\section{Parameter estimation}\label{sec:method}

The parameter estimation aims to extract the $g$-mode frequencies and solve the universal relations. The following subsections provide detailed descriptions of this procedure.

\subsection{Signal reconstruction}\label{subsec:reconst}

\begin{figure}[b]
    \centering
    \includegraphics[width=\columnwidth]{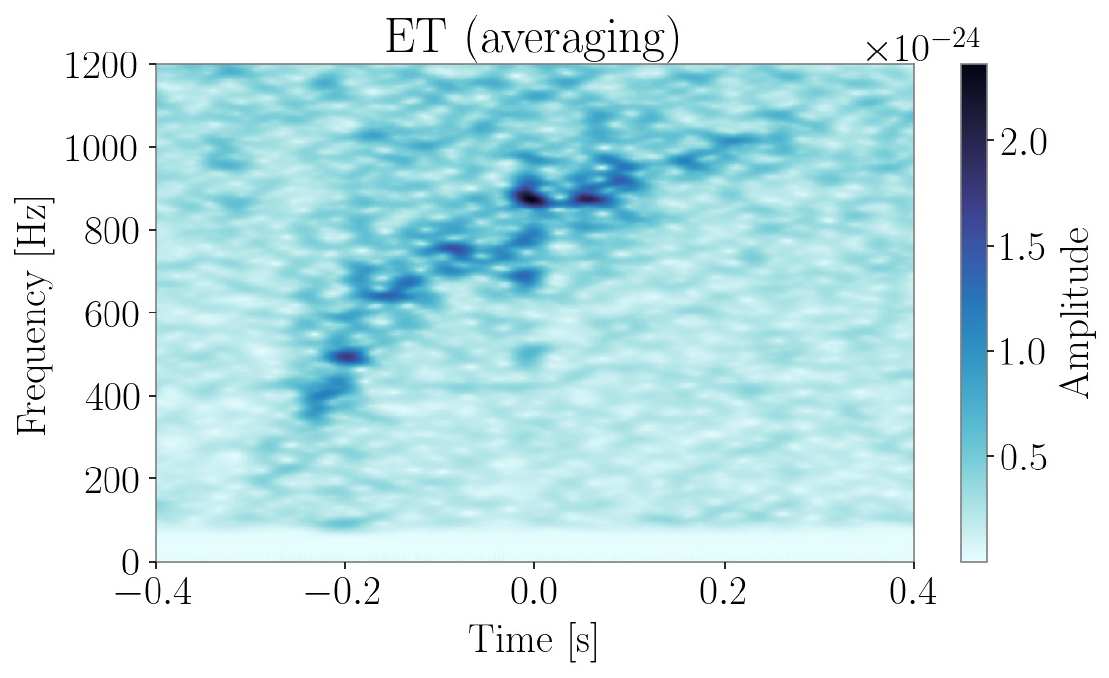}
    \caption{Averaged spectrogram over the three ET configurations, each combining the same he3.5 waveform data with their respective detector noise.}
    \label{fig:stft_he35_17kpc_noise}
\end{figure}

Both experimental data and noise-added simulated data undergo the same reconstruction algorithm to extract GW transients. For this purpose, we employ the cWB algorithm~\cite{ref:Klimenko2008, ref:Klimenko2016, ref:Drago2021}, an excess-power algorithm designed to detect and reconstruct GWs with minimal assumptions on waveform morphologies. This reconstruction step is applied to the raw data, illustrated by the spectrogram Fig.~\ref{fig:stft_he35_17kpc_noise}, to separate GW signal from detector noise. In this figure, he3.5 model waveform is generated from a source distance of 17~kpc. Following this, the HHT procedure to extract $g$-mode information from CCSNe data.

We used PycWB~\cite{ref:Xu2024}, a modularized Python package for the cWB reconstruction. We adopted the internal parameters and thresholds from the LIGO-Virgo targeted search for CCSNe during the O1 and O2 runs~\cite{ref:Abbott2020_ccsn}. The procedure for cWB reconstruction involves the multiple steps. First, time-series data is whitened and turned into a time-frequency domain using the Wilson-Daubechies-Meyer wavelet transform~\cite{ref:Necula2012} for several different time-frequency resolutions. Pixel energy is maximized over all sky locations, and pixels exceeding a specified threshold are retained for each time-frequency map. Subsequently, neighboring pixels are clustered as coincident pixels in the multi-resolution time-frequency maps, and coherent clusters across multiple detectors serve as triggers. Each trigger undergoes evaluation using the constrained maximum likelihood method~\cite{ref:Klimenko2008}. For accepted events, waveforms are reconstructed by applying the inverse wavelet transform to the selected pixels.

\begin{figure}[t]
    \centering
    \includegraphics[width=\columnwidth]{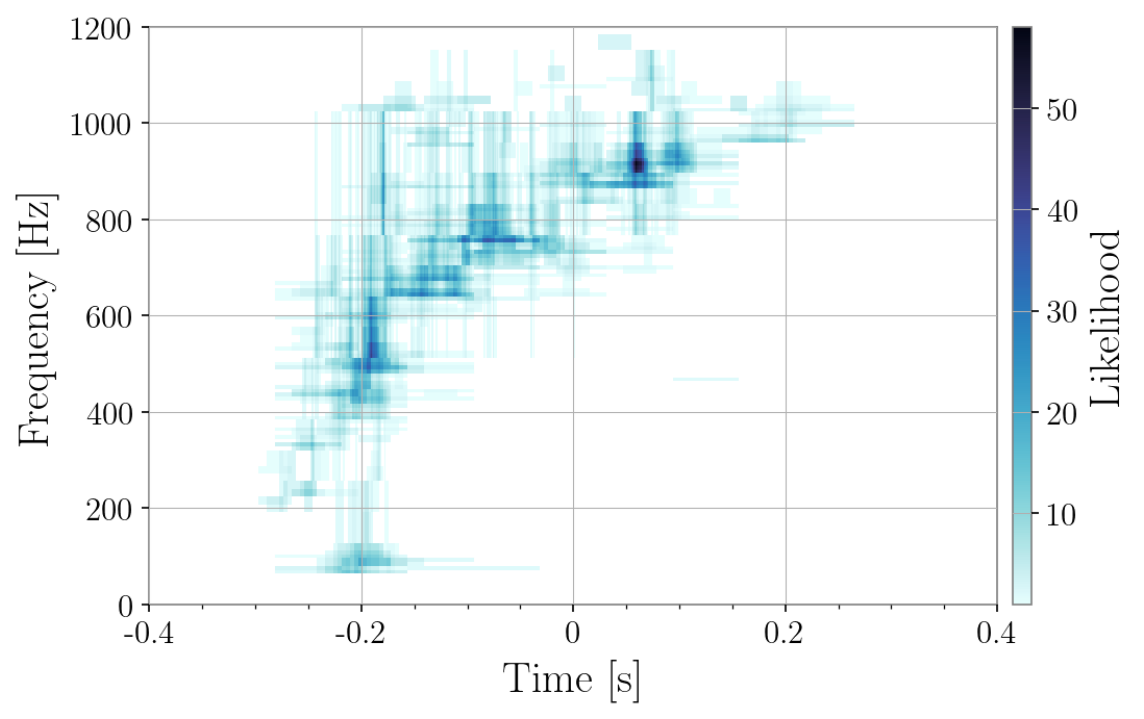}
    \caption{Pixel distribution of the he3.5 sample after cWB reconstruction.}
        \label{fig:he35_17kpc_noise_cwb}
\end{figure}
\begin{figure}[t]
\includegraphics[width=\columnwidth]{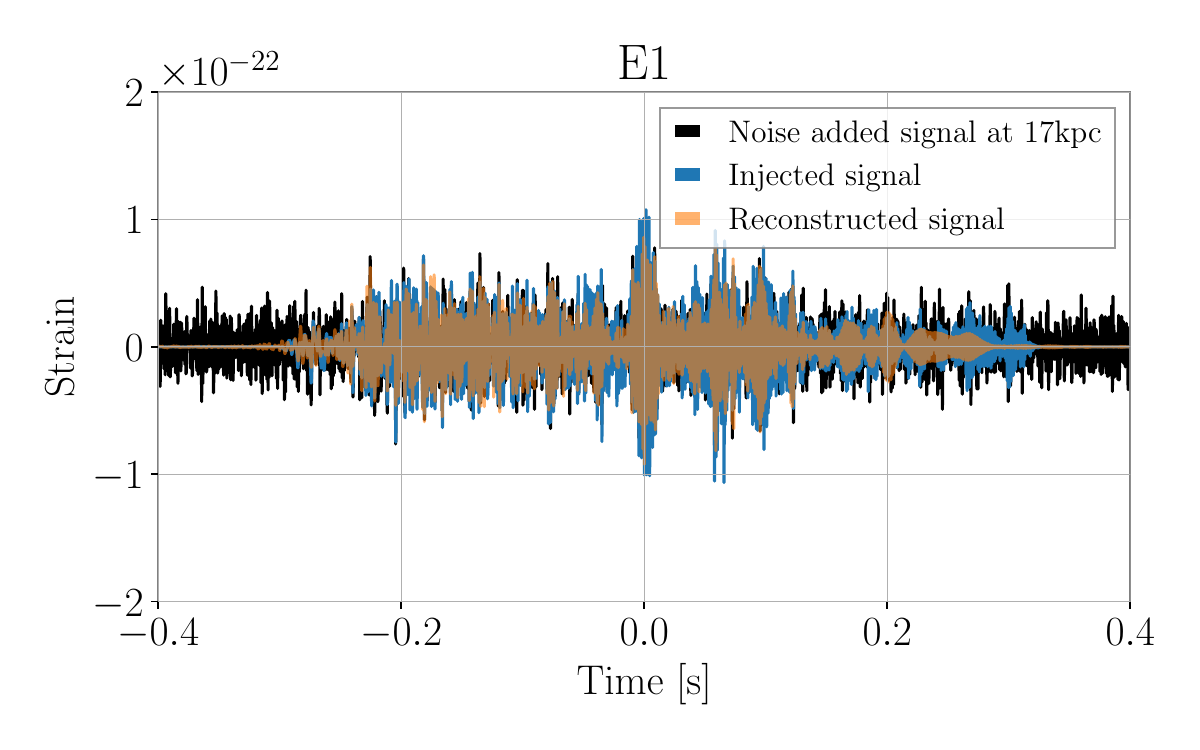}
    \caption{Reconstructed signal from cWB algorithm (orange) obtained after processing a combination of he3.5 model and ET1 detector noise.}
    \label{fig:he35_17kpc_noise}
\end{figure}

Eventually, selected pixels by the cWB algorithm are shown in Fig.~\ref{fig:he35_17kpc_noise_cwb}. From such likelihood distribution, we obtain the reconstructed signal from the ET1 detector as shown in Fig.~\ref{fig:he35_17kpc_noise}. Same reconstruction procedure applies to the y20 model and shown similar results.

For signals injected into detector noise, we can measure how well the reconstructed waveforms match the original waveform by making use of the \textit{network match}~\cite{ref:abbott2016}. Let $w_k$ represent the reconstructed waveform and $h_k$ be the injected waveform. In this context, the \textit{network match} denoted $\mathcal{N}$ is defined by the following equation:
\begin{equation}
    \mathcal{N} = \frac{\displaystyle \sum_{k=1}^{K}(w_k|h_k)}{\sqrt{\displaystyle \sum_{k=1}^{K}(w_k|w_k)}\sqrt{\displaystyle \sum_{k=1}^{K}(h_k|h_k)}},
    \label{eq:network_match}
\end{equation}
where $K$ is the number of detectors, i.e., $K=3$ in this analysis, and $(\cdot|\cdot)$ denotes a noise-weighted inner product defined using the one-sided power spectral density $S(f)$ of the detector noise, which is defined as
\begin{equation}
    (a|b) = 4 \mathrm{Re}\int_{f_{\mathrm{min}}}^{f_{\mathrm{max}} }\frac{\tilde{a}(f)\tilde{b}^*(f)}{S(f)}\mathrm{d}f.
\end{equation}
Here, $f_{\mathrm{min}}$ and $f_{\mathrm{max}}$ represent the integration ranges of the Fourier transform, which are set to $100$ Hz and $1200$ Hz, respectively.
The network match takes a value between $0$ and $1$. If $\mathcal{N} = 1$, the reconstructed waveforms perfectly match the injected waveforms, and $\mathcal{N} =  0$ indicates no match. To evaluate how detectable the injected signal $h_k$ is under the detector sensitivity $S(f)$, we define the network signal-to-noise ratio (SNR) as
\begin{equation}
   \rho_{\text{network}} = \sqrt{\sum_{k=1}^K (h_k|h_k)}.
   \label{eq:network_snr}
\end{equation}

\subsection{Hilbert-Huang transform}\label{subsec:hht}
The HHT~\cite{ref:Huang1998} comprises two steps: empirical mode decomposition (EMD) and Hilbert spectral analysis.
In the first step, EMD decomposes a time-series signal into a sum of intrinsic mode functions (IMFs) denoted $c_k(t)$, where $k$ represents the $k$th IMF, alongside a residual component $r(t)$. Therefore, a given input signal $s(t)$ can be decomposed as follows:
\begin{equation}
  s(t) = c_{1..N}(t) + r(t),
  \label{eq:signal-imf}
\end{equation}
where $N$ is the order of empirical mode decomposition and the new notation $c_{i..j}(t)$ represents a simplified form of the expression:
\begin{equation}
c_{i..j}(t) = \sum_{k=i}^{j} c_{k}(t)
\end{equation}
Each IMF, $c_k(t)$, represents an oscillatory mode found in the original signal, while $r(t)$ accounts for a residual trend not captured by the EMD.
Following EMD, the subsequent step is Hilbert spectral analysis, which employs the Hilbert transform to compute each IMF's, or any combination of IMF's, instantaneous amplitude (IA) and frequency (IF).
This is usually achieved by computing the analytic signal $\tilde{c}_k(t)$ from the $k$th IMF $c_k(t)$, as following
\begin{equation}
    \tilde{c}_k(t) = c_k(t) + i\mathcal{H}[c_k(t)],
\end{equation}
where $\mathcal{H}[c_k(t)]$ denotes the Hilbert transform of $c_k(t)$, defined by
\begin{equation}
    \mathcal{H}[c_k(t)] = \frac{1}{\pi}\mathrm{PV}\int_{-\infty}^{\infty} \frac{c_k(t')}{t-t'}\mathrm{d}t'.
\end{equation}
Here, PV denotes the Cauchy principal value. The instantaneous amplitude $\mathrm{IA}_k(t)$ and the instantaneous frequency $\mathrm{IF}_k(t)$ derived from the analytic signal $\tilde{c}_i(t)$ are defined 
respectively, as
\begin{equation}
    \begin{aligned}
    \mathrm{IA}_k(t) &= \abs{\tilde{c}_k(t)}, \\
    \mathrm{IF}_k(t) &= \frac{1}{2\pi} \frac{\mathrm{d}}{\mathrm{d}t}\mathrm{arg}\left[\tilde{c}_k(t)\right],
    \end{aligned}
\end{equation}

The initial HHT uses a basic EMD as a decomposition method but often encounters problems such as mode mixing and mode splitting~\cite{ref:Wu2009}.
These issues were addressed in a series of empirical updates over the years aiming to improve stability.
\begin{itemize}[leftmargin=*]
\item\citeauthor{ref:Wu2009}~\cite{ref:Wu2009} proposed an extension of EMD called Ensemble EMD (EEMD). EEMD works by adding white Gaussian noise to the original signal and performing EMD on multiple realizations of the noisy signal. By averaging the IMFs obtained from different realizations, EEMD improves the accuracy of the decomposition.
\item Another extension called Complementary EEMD (CEEMD) was proposed by \citeauthor{ref:Yeh2010}~\cite{ref:Yeh2010} to enhance the decomposition further. In this method, white Gaussian noise is added in pairs---both positive and negative---to the original signal, producing two sets of ensemble IMFs. The final IMFs are the ensemble of both the IMFs with positive and negative noises. The algorithmic steps of this technique are described in Appendix~\ref{app:ceemd}. 
\end{itemize}

Extensive details on the computation of the HHT are provided in Refs.~\cite{ref:Kaneyama2016, ref:Yoda2023,ref:Takeda2021}. 

Moreover, the detailed parametrization of the HHT used in this analysis is listed in Table~\ref{tab:hht_parameters} and applies to both the he3.5 and y20 models.

\begin{table}[h]
    \caption{Parameters of CEEMD.}
    \label{tab:hht_parameters}
    \centering
    \begin{ruledtabular}
    \begin{tabular}{l c c}
        Total number of IMFs & 4 \\
        Number of ensemble for CEEMD& 1000 \\
        Standard deviation of added Gaussian noise & 0.8\\
        Stoppage criterion & $7 \times 10^{-4}$ \\
    \end{tabular}
    \end{ruledtabular}
\end{table}

\subsection{Solving universal relation}\label{subsec:u-relation}
Inference of the physical properties of PNSs involves utilizing the $g$-mode universal relations~\cite{ref:Torres2019, ref:Torres2021}:
\begin{equation}
    \label{eq:universal_gmode}
    f(t) = a + bx(t) + cx(t)^2 + dx(t)^3,
\end{equation}
where $f(t)$ is the frequency of the $g$-mode, and $x(t)$ represents the PNS mass $M_{\mathrm{PNS}}$ in solar masses divided by the squared radius $R_{\mathrm{PNS}}$ in kilometers. Using the computed IFs and the parameters in Table~\ref{tab:gmode_coefficients}, we solve the cubic equation Eq.~\eqref{eq:universal_gmode} for each time to obtain the time evolution of $M_{\mathrm{PNS}}/R_{\mathrm{PNS}}^2$. Solutions to the cubic equations are computed via the eigenvalues of the companion matrix~\cite{ref:LAPACK}.
Finally, we average the estimated values from the three ET detectors.

\begin{table}[t]
    \caption{$g$-mode coefficients of the universal relations with $x = M_{\mathrm{PNS}}/R_{\mathrm{PNS}}^2$~\cite{ref:Torres2019, ref:Torres2021}.}
    \label{tab:gmode_coefficients}
    \centering
    \renewcommand{\arraystretch}{1.3}
    \begin{ruledtabular}
    \begin{tabular}{c c c c c}
         mode & $a$ & $b/10^5$ & $c/10^6$ & $d/10^9$ \\
        \hline
        $^2g_1$ & 0 & 8.67 & $-51.9$ & 0 \\
        $^2g_2$ & 0 & 5.88 & $-86.2$ & 4.67 \\
    \end{tabular}
    \end{ruledtabular}
\end{table}

Eventually, to compare method outcomes in the discussion section, we utilize the root mean squared error (RMSE), defined by
\begin{equation}
    \mathrm{RMSE} = \sqrt{\frac{1}{N}\sum_i (y_i - \hat{y}_i)^2},
    \label{eq:rmse}
\end{equation}
where $N$ is the total number of samples, $y_i$ represents the true values, and $\hat{y}_i$ is the estimated one. 

\section{Results and discussion}\label{sec:result}

In this section, we first present our results for the pure GW signals. 
Next, we will discuss the results for GW signals injected into ET detector noise. For each detector and distance, we generated a hundred randomized sequences of colored Gaussian noise. The distances considered for the GW signals range from 1~kpc to 29~kpc.

\begin{figure}[h]
    \centering
        \centering
        \includegraphics[width=\columnwidth]{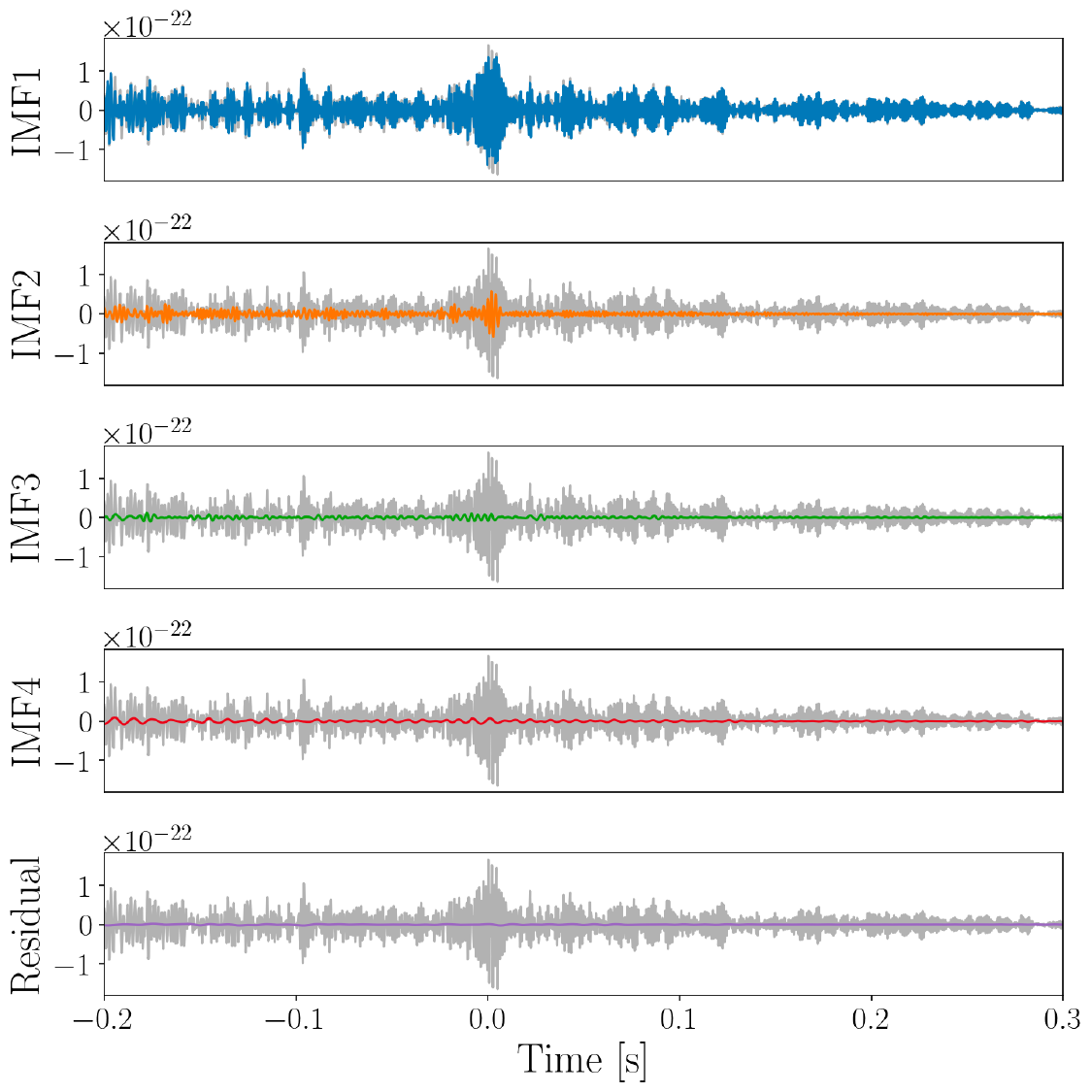}
    \caption{IMFs and residual of the HHT for the pure he3.5 signal from 10 kpc, observed at ET1 detector. The gray waveform in each plot represents the original signal.}
    \label{fig:hht_pure_he35_10kpc_imfs}
\end{figure}

\begin{figure*}[htb]
    \centering
    \begin{tabular}{cc}
      \begin{minipage}[t]{0.49\textwidth}
        \centering
        \includegraphics[width=\textwidth]{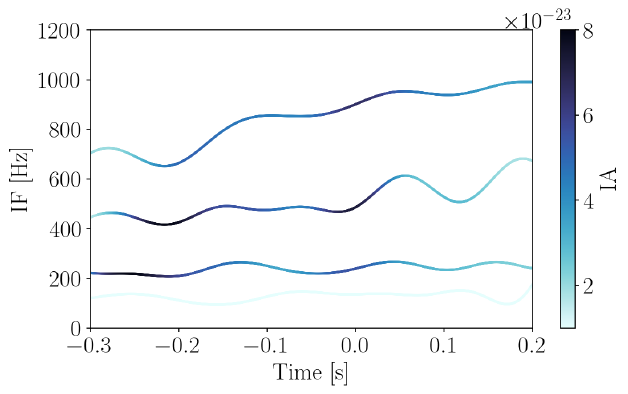}
        \subcaption{HHT}
        \label{fig:pure_he35_10kpc_hht}
      \end{minipage} &
        \begin{minipage}[t]{0.49\textwidth}
        \centering
        \includegraphics[width=\textwidth]{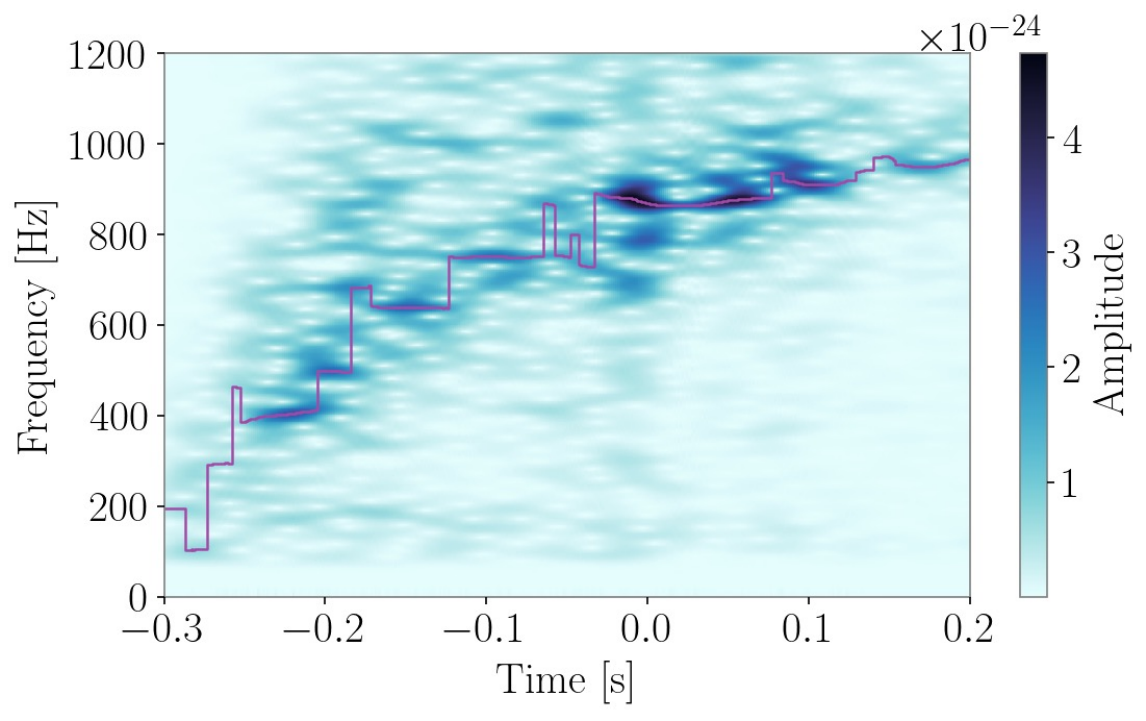}
        \subcaption{STFT}
        \label{fig:pure_he35_10kpc_stft}
      \end{minipage}
    \end{tabular}
    \caption{Time-frequency maps of the pure he3.5 signal from 10 kpc, observed at ET1 detector.}
\end{figure*}

\subsection{Simple case: Pure gravitational wave signals}\label{subsec:pure_signal}
In this first step, we apply our approach to the pure GW signals, which were discussed in Sec.~\ref{subsec:waveform}. In this case, cWB reconstruction is not required.

Figure~\ref{fig:hht_pure_he35_10kpc_imfs} shows the extracted IMFs of the he3.5 signal using CEEMD with parameters listed in Table~\ref{tab:hht_parameters}. We then apply the Hilbert spectral analysis to estimate the IAs and IFs from the obtained IMFs. The resulting time-frequency map is shown in Fig.~\ref{fig:pure_he35_10kpc_hht}. In this plot, the frequencies of four IMFs are shown, with IMF1, IMF2, IMF3, and IMF4 representing the highest to the lowest frequencies.
Figure~\ref{fig:pure_he35_10kpc_stft} shows the time-frequency map generated using STFT for comparison. Purple line in the STFT time-frequency map corresponds to the frequency with the maximum amplitude for each time bin. 
We compare the IFs of the HHT map and the purple line of STFT map to estimate the frequency of the $g$-mode.

\begin{figure}[h]
    \centering
        \centering
        \includegraphics[width=\columnwidth]{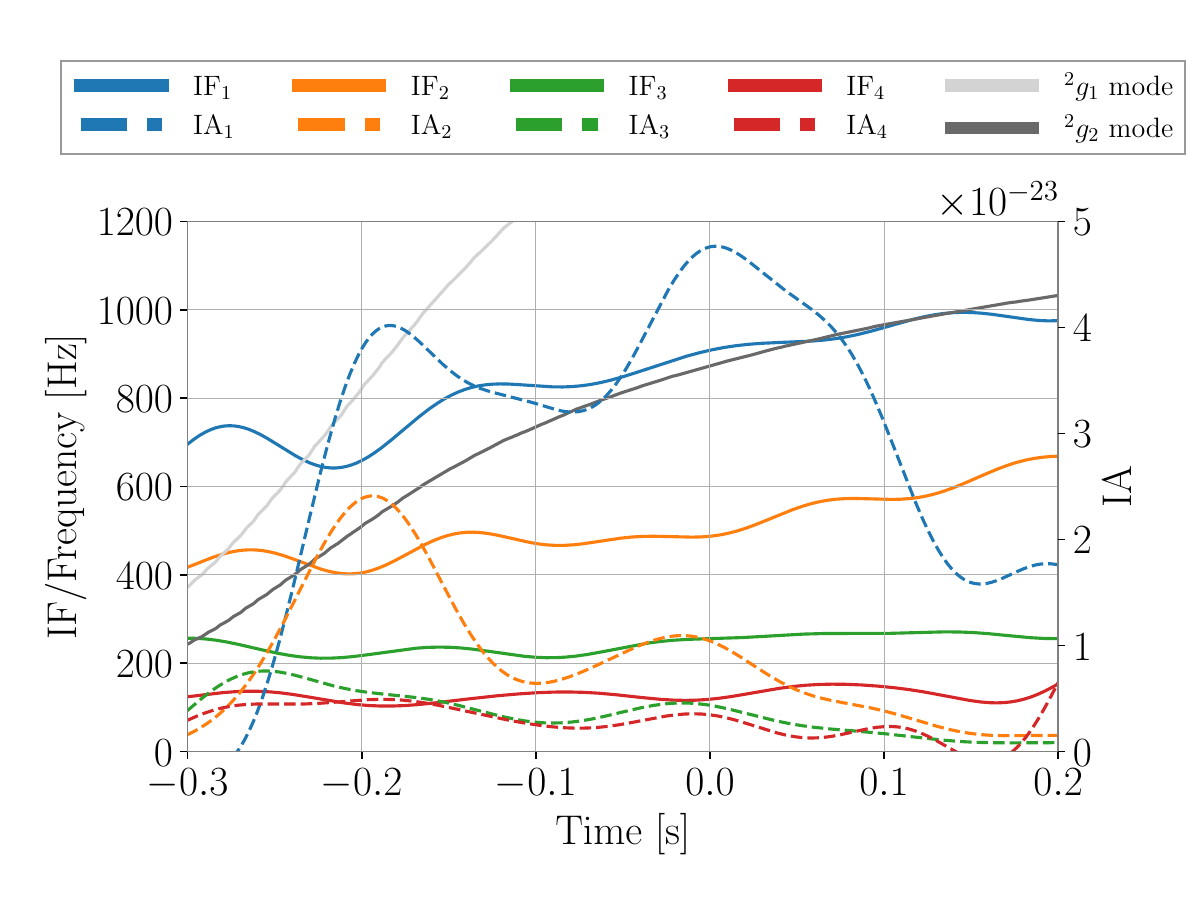}
    \caption{IFs and IAs of the HHT from pure signal for each model. The $^2g_1$ and $^2g_2$ modes from the simulations are also plotted as the reference.}
    \label{fig:hht_10kpc_he35_g1g2}
\end{figure}

\begin{figure}[h]
    \centering
        \centering
        \includegraphics[width=\columnwidth]{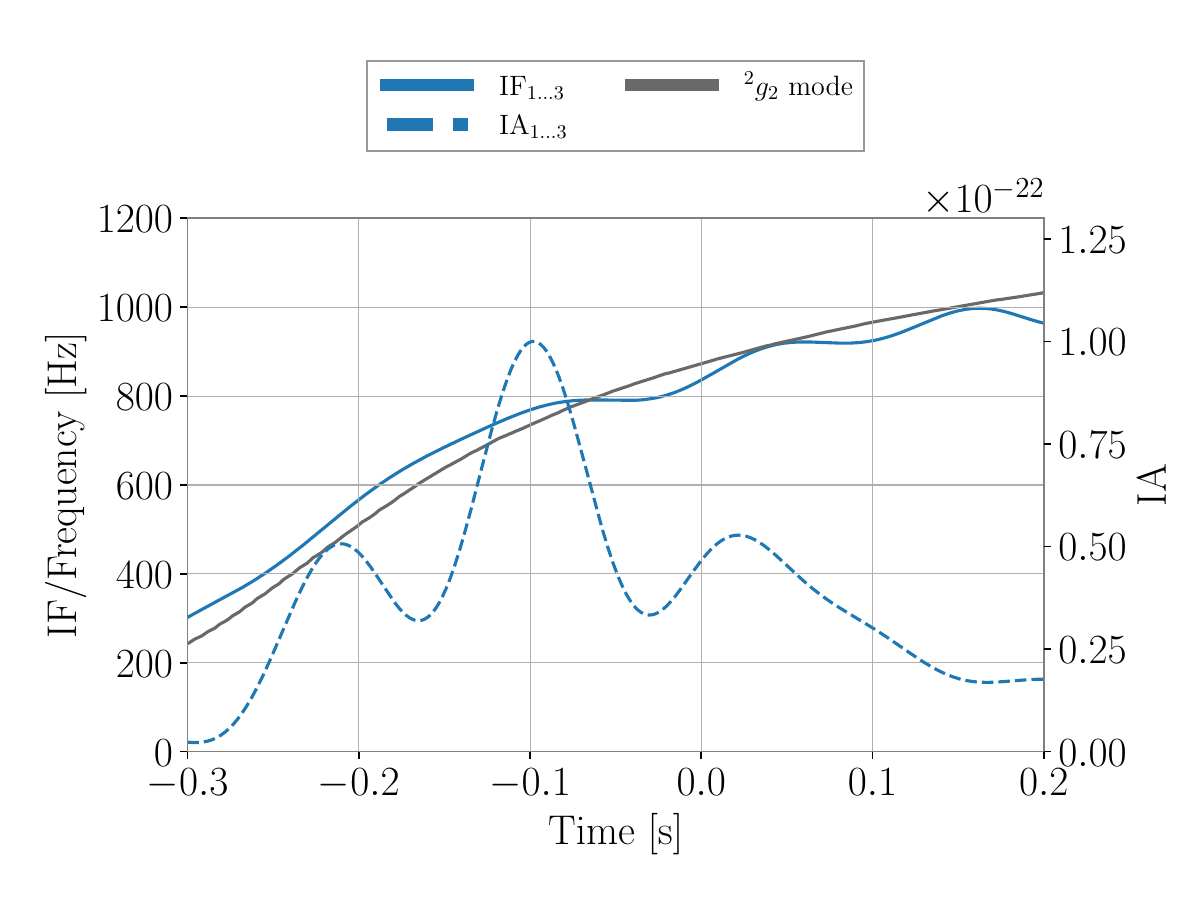}
    \caption{$\mathrm{IF}_{1..3}$ and $\mathrm{IA}_{1..3}$ from the HHT. The $^2g_2$ mode from the simulations is plotted as the reference.}
    \label{fig:hht_if123_he35_10kpc}
\end{figure}

Figure~\ref{fig:hht_10kpc_he35_g1g2} shows the IFs and IAs of the HHT along with the $^2g_1$ and $^2g_2$ modes associated with the he3.5 model. The $\mathrm{IF}_{1}$ appears to estimate the intermediate frequency between the $^2g_1$ and $^2g_2$ modes from $-0.2$~s to $-0.1$~s. In general, the evolution of $g$-modes spans a wide frequency range, from a low frequency of $\sim 200$~Hz to several thousand Hz. This wide frequency range makes it challenging to decompose the signal into a single IMF, potentially leading to the mode splitting issue~\cite{ref:Wu2009}. Our result also indicates that this problem occurs, as the $^2g_2$ mode appears to be separated into several IFs, such as $\mathrm{IF}_{1}$ around 0 s, $\mathrm{IF}_{2}$ around $-0.2$~s, and $\mathrm{IF}_{3}$ around $-0.3$~s. In this case, summing the IMFs before applying Hilbert spectral analysis is one empirical method to mitigate the effect of mode splitting.
Figure~\ref{fig:hht_if123_he35_10kpc} shows the sums of the three IFs and three IAs, labeled as $\mathrm{IF}_{1..3}$ and $\mathrm{IA}_{1..3}$, respectively. Note that $\mathrm{IMF}_{4}$ is excluded from the summation due to its relatively small IA and the estimated IF being below 200 Hz, which is outside the frequency range of the $^2g_2$ mode.

To estimate $M_{\mathrm{PNS}}/R_{\mathrm{PNS}}^2$, the extracted frequencies were averaged across the three ET detectors, and the universal relation for the $^2g_2$ mode, as given in Eq.~\eqref{eq:universal_gmode}, was solved at each time step.
Figure~\ref{fig:pure_urs} shows the calculated $M_{\mathrm{PNS}}/R_{\mathrm{PNS}}^2$ for both the he3.5 and y20 models, using the extracted frequencies from the STFT map [Fig.~\ref{fig:pure_he35_10kpc_stft}] and $\mathrm{IF}_{1..3}$. The green lines in Fig.~\ref{fig:pure_urs} represent the true values obtained from the numerical simulations and are used to calculate the RMSEs.
In Fig.~\ref{fig:pure_he35_10kpc_stft}, the he3.5 estimate exhibits time discontinuities inherent to the STFT as one of its limitations. In this case, neglecting this systematic uncertainty, STFT method still reasonably estimates the continuous ratio $M_{\mathrm{PNS}}/R_{\mathrm{PNS}}^2$ across all times. Therefore, we will use it as reference to compare with the new HHT method.
In contrast, the estimates from the HHT are smoother and align more closely with the true values.
The ratio of the RMSE for the HHT to that for STFT is 0.66, indicating that the HHT provides approximately 33.9\% lower error compared to STFT.

For the y20 model, the RMSE of the HHT is approximately 4.1\% lower than that of STFT. Both methods yield to RMSE larger values compared to the he3.5 model, as the estimates progressively deviate from the $^2g_2$ mode after 0 s. 
The $g$-mode estimated from the universal relation is expected to exhibit greater uncertainties at higher frequencies, as discussed by Bizouard~\textit{et al.}~\cite{ref:bizouard2021inference} and Sotani~\textit{et al.}~\cite{ref:sotani2024universality}, both of whom also use the same empirical formula, Eq.~\eqref{eq:universal_gmode}, originally derived by Torres \textit{et al.}~\cite{ref:Torres2019, ref:Torres2021}.
Despite this, both methods provide similar estimates, indicating that both the HHT and STFT can extract the frequencies of the $^2g_2$ mode with comparable accuracy.

\subsection{Detector noise-added gravitational wave signal} \label{subsec:noise_signal}
The GW waveforms were injected in simulated 100 different detector noises and the injected waveforms in each noise were reconstructed by cWB~\cite{ref:Klimenko2008, ref:Klimenko2016, ref:Drago2021}.

\begin{figure*}[htb]
    \centering

    \includegraphics[width=0.95\textwidth]{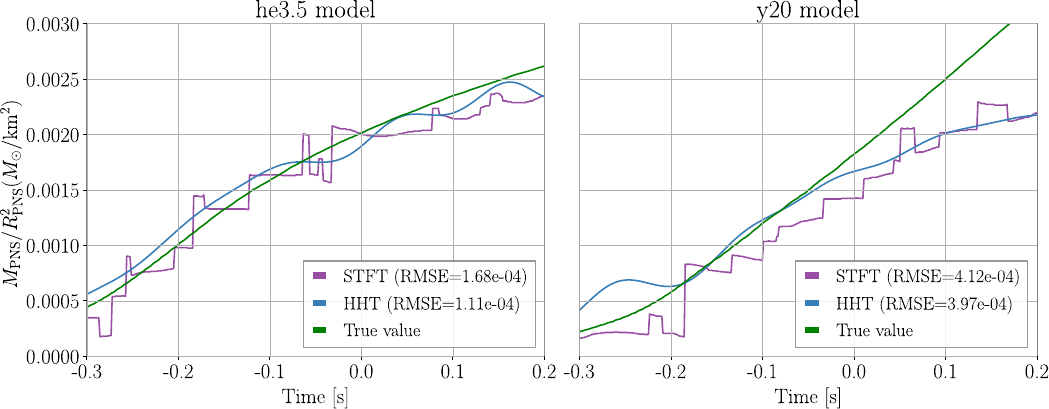}
    \caption{Estimates of $M_{\mathrm{PNS}}/R_{\mathrm{PNS}}^2$ calculated using the pure GW signal from 10 kpc. True values are obtained from numerical simulation.}
    \label{fig:pure_urs}
    \hfill\hfill
    
    \includegraphics[width=0.98\textwidth]{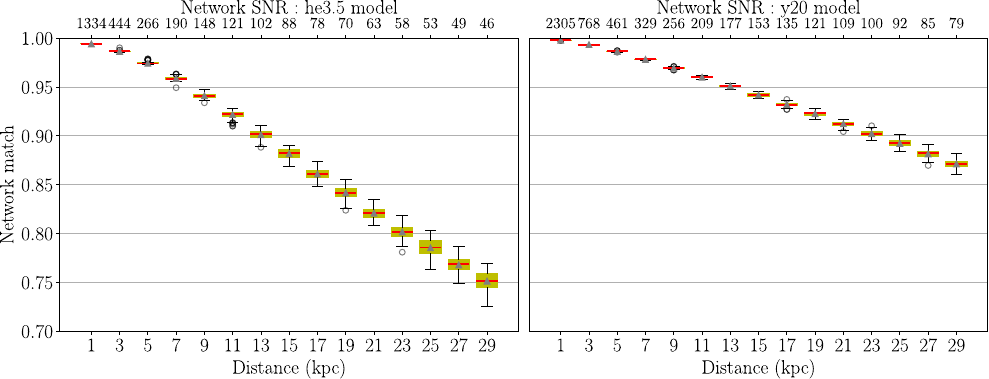}
    \caption{Distributions of the network match as the function of distance or network SNR. Each box plot shows its an interquartile range for 100 different noise realizations.}
    \label{fig:network_match_cwb}
    \hfill\hfill
    
    \includegraphics[width=0.98\textwidth]{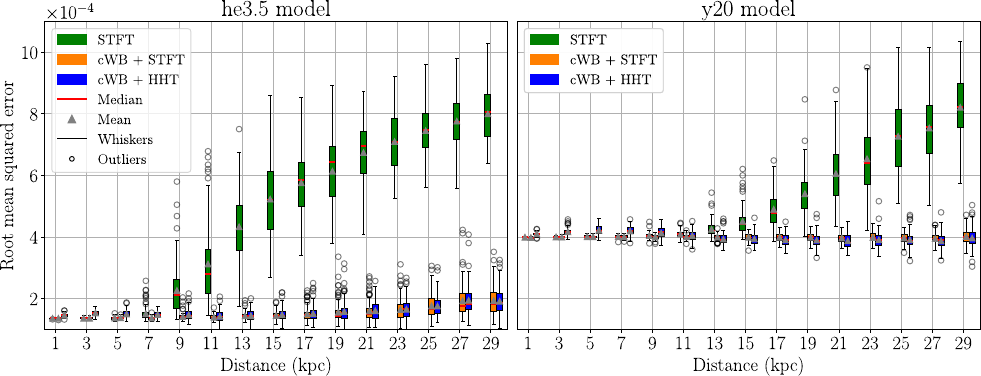}
    \caption{Distributions of the RMSE of $M_{\mathrm{PNS}}/R_{\mathrm{PNS}}^2$ as a function of source distance for the three methods. Each box plot shows its an interquartile range for 100 different noise realizations.}
    \label{fig:rmse}
\end{figure*}
Figure~\ref{fig:network_match_cwb} shows the distributions of the network match defined in Eq.~\eqref{eq:network_match} as the function of each distance or network SNR in Eq.~\eqref{eq:network_snr} for each signal model. 
The plot exhibits the expected behavior; when the source distance is small, the cWB can accurately reconstruct the waveform. As the distance increases, the network match decreases, and its variance increases. Similar results are observed for the y20 case, despite that the network match is larger than the he3.5 case.

Estimation of $M_{\mathrm{PNS}}/R_{\mathrm{PNS}}^2$ is conducted via three methods; (i) the method STFT denotes the extraction of frequencies from the original STFT map using the frequencies with maximum amplitude for each time bin, (ii) cWB + STFT refers to the application of the cWB waveform reconstruction and the extraction of frequencies from the reconstructed STFT map, and (iii) cWB + HHT denotes the use of the cWB waveform reconstruction and the HHT.

The RMSE for each method is shown in Fig.~\ref{fig:rmse} for each signal model. In the he3.5 case, for signals with distances less than 7 kpc, the three methods exhibit almost the same RMSE. However, as the source distance increases, the STFT-based method experiences significantly increased RMSEs, due to the influence of noise other than the $^2g_2$ mode signal. As the distance increases, the two methods employing the cWB waveform reconstruction reduce the effect of noise through the extraction of coherent $^2g_2$ mode signals across multiple detectors. Regarding the y20 case, the RMSE of the STFT-based method increases at the distances greater than 15 kpc, while for the he3.5 case, the RMSE starts to increase at 9 kpc. As for the accuracy of signal reconstruction shown in Fig.~\ref{fig:network_match_cwb}, the network match of the y20 signal at 15 kpc is equal to that of the he3.5 signal at 9 kpc. These results imply that for signals with the same reconstruction accuracy, the estimation error of $M_{\mathrm{PNS}}/R_{\mathrm{PNS}}^2$ is also similar. Furthermore, the two methods, cWB + STFT and cWB + HHT, have equivalent RMSE values and the estimation. Based on these results, we conclude that both the cWB + STFT and cWB + HHT methods provide estimations of $M_{\mathrm{PNS}}/R_{\mathrm{PNS}}^2$ with comparable accuracy.

In conclusion, our HHT approach showed similar results to the STFT-based one in terms of the RMSE, highlighting the potential of this approach as a complementary method to extract the physical properties of PNSs from GW signals of CCSNe.

\section{Conclusions\label{sec:concl}}
In this study, we demonstrated the use of the HHT for estimating $M_{\mathrm{PNS}}/R_{\mathrm{PNS}}^2$ of PNSs from their $g$-mode GW signals, using both the he3.5 and y20 signals. We first applied it to the pure GW signal and confirmed the effectiveness of our method to extract the $g$-mode frequencies. We then tested our approach, the combination of the cWB reconstruction and the HHT, for the signal injected into the ET detector noise. Our new approach showed comparable results to the STFT-based one in terms of the RMSE.

For the extraction of the $^2g_2$ mode using the HHT, the multiple $g$-modes appear in the IMFs. In this study, we utilize the combined IMFs to reduce the effect of mode splitting. Further improved techniques for addressing mode mixing and splitting will enhance the performance of extraction. Furthermore, such techniques can be applied not only to the extraction of multiple $g$-modes but also to the extraction of other features such as the SASI~\cite{ref:Roberts2012, ref:Kuroda2016}.
Our new complementary approach shows comparable results, and having two independent methods that yield equivalent accuracy might be valuable in real-world applications where the actual value is unknown. To validate the efficacy of our method further, testing with additional simulation data is necessary. For waveform reconstruction, we utilized the cWB algorithm, which does not assume a waveform model in advance. We will also use another reconstruction method, BayesWave~\cite{ref:Cornish2015, ref:Cornish2021}, and test the effectiveness of our method.

\begin{acknowledgments}
The authors would like to thank Jade Powell and Bernhard M\"uller for providing their simulation data.
This research was supported in part by the Japan Society for the Promotion of Science (JSPS) Grant-in-Aid for Scientific Research [Nos. 22H01228, 23K22499 (K.\ Somiya), 21K13926 (K.\ Sakai), 22K03614 (K.\ Oohara) and Nos. 19H01901, 23H01176, 23K25872 and 23H04520 (H.\ Takahashi)] and Grant-in-Aid for JSPS Fellows [No. 22KF0329 (M.\ Meyer-Conde)].
This research was supported by the Joint Research Program of the Institute for Cosmic Ray Research, University of Tokyo, and Tokyo City University Prioritised Studies.
\end{acknowledgments}
\appendix
\section{COMPLEMENTARY ENSEMBLE EMPIRICAL MODE DECOMPOSITION}\label{app:ceemd}
We employ the complementary ensemble empirical mode decomposition~\cite{ref:Yeh2010} to decompose signals into intrinsic mode functions. The decomposition process of a signal $s(t)$ is as follows:
\begin{enumerate}
    \item Generate $N$ time-series sequences $\{n_k(t)\}\ (k=1, \dots, N)$, where each sequence represents white Gaussian noise with mean zero and standard deviation $\sigma$.
    \item Construct noise-added signal $s_k^{(+)}(t)$ and noise-subtracted signal $s_k^{(-)}(t)$ to the target signal $s(t)$:
    \begin{equation}
      s_k^{(\pm)}(t) = s(t) \pm n_k(t),
    \end{equation}
    for $k=1, \dots, N$.
    \item Employ EMD to each $s_k^{(\pm)}(t)$ with the stoppage criterion based on a Cauchy convergence test with a predetermined value $\epsilon$~\cite{ref:Kaneyama2016} to obtain IMFs $\{c_{i,k}^{(\pm)}(t)\}$, alongside the residuals $r_k^{(\pm)}(t)$:
    \begin{equation}
      s_k^{(\pm)}(t) = \sum_i c^{(\pm)}_{i, k}(t) + r_k^{(\pm)}(t).
    \end{equation}
    \item Calculate the final IMFs and residue by
    \begin{align}
      c_i(t) &= \frac{1}{2N}\sum_{k=1}^N \pqty{c^{(+)}_{i, k}(t)+c^{(-)}_{i, k}(t)},\\
      r(t) &= \frac{1}{2N}\sum_{k=1}^N \pqty{r^{(+)}_k(t)+r^{(-)}_k(t)}.
    \end{align}
\end{enumerate}

\clearpage 
\bibliography{apssamp}

\begin{thebibliography}{66}%
\makeatletter
\providecommand \@ifxundefined [1]{%
 \@ifx{#1\undefined}
}%
\providecommand \@ifnum [1]{%
 \ifnum #1\expandafter \@firstoftwo
 \else \expandafter \@secondoftwo
 \fi
}%
\providecommand \@ifx [1]{%
 \ifx #1\expandafter \@firstoftwo
 \else \expandafter \@secondoftwo
 \fi
}%
\providecommand \natexlab [1]{#1}%
\providecommand \enquote  [1]{``#1''}%
\providecommand \bibnamefont  [1]{#1}%
\providecommand \bibfnamefont [1]{#1}%
\providecommand \citenamefont [1]{#1}%
\providecommand \href@noop [0]{\@secondoftwo}%
\providecommand \href [0]{\begingroup \@sanitize@url \@href}%
\providecommand \@href[1]{\@@startlink{#1}\@@href}%
\providecommand \@@href[1]{\endgroup#1\@@endlink}%
\providecommand \@sanitize@url [0]{\catcode `\\12\catcode `\$12\catcode `\&12\catcode `\#12\catcode `\^12\catcode `\_12\catcode `\%12\relax}%
\providecommand \@@startlink[1]{}%
\providecommand \@@endlink[0]{}%
\providecommand \url  [0]{\begingroup\@sanitize@url \@url }%
\providecommand \@url [1]{\endgroup\@href {#1}{\urlprefix }}%
\providecommand \urlprefix  [0]{URL }%
\providecommand \Eprint [0]{\href }%
\providecommand \doibase [0]{https://doi.org/}%
\providecommand \selectlanguage [0]{\@gobble}%
\providecommand \bibinfo  [0]{\@secondoftwo}%
\providecommand \bibfield  [0]{\@secondoftwo}%
\providecommand \translation [1]{[#1]}%
\providecommand \BibitemOpen [0]{}%
\providecommand \bibitemStop [0]{}%
\providecommand \bibitemNoStop [0]{.\EOS\space}%
\providecommand \EOS [0]{\spacefactor3000\relax}%
\providecommand \BibitemShut  [1]{\csname bibitem#1\endcsname}%
\let\auto@bib@innerbib\@empty
\bibitem [{\citenamefont {Abbott}\ \emph {et~al.}(2019)\citenamefont {Abbott} \emph {et~al.}}]{ref:Abbott2019}%
  \BibitemOpen
  \bibfield  {author} {\bibinfo {author} {\bibfnamefont {B.~P.}\ \bibnamefont {Abbott}} \emph {et~al.} (\bibinfo {collaboration} {LIGO Scientific Collaboration and Virgo Collaboration}),\ }\bibfield  {title} {\bibinfo {title} {{GWTC-1}: A gravitational-wave transient catalog of compact binary mergers observed by {LIGO} and {Virgo} during the first and second observing runs},\ }\href {https://doi.org/10.1103/PhysRevX.9.031040} {\bibfield  {journal} {\bibinfo  {journal} {Phys. Rev. X}\ }\textbf {\bibinfo {volume} {9}},\ \bibinfo {pages} {031040} (\bibinfo {year} {2019})}\BibitemShut {NoStop}%
\bibitem [{\citenamefont {Abbott}\ \emph {et~al.}(2021)\citenamefont {Abbott} \emph {et~al.}}]{ref:Abbott2020}%
  \BibitemOpen
  \bibfield  {author} {\bibinfo {author} {\bibfnamefont {R.}~\bibnamefont {Abbott}} \emph {et~al.} (\bibinfo {collaboration} {LIGO Scientific Collaboration and Virgo Collaboration}),\ }\bibfield  {title} {\bibinfo {title} {{GWTC-2}: Compact binary coalescences observed by {LIGO} and {Virgo} during the first half of the third observing run},\ }\href {https://doi.org/10.1103/PhysRevX.11.021053} {\bibfield  {journal} {\bibinfo  {journal} {Phys. Rev. X}\ }\textbf {\bibinfo {volume} {11}},\ \bibinfo {pages} {021053} (\bibinfo {year} {2021})}\BibitemShut {NoStop}%
\bibitem [{\citenamefont {Abbott}\ \emph {et~al.}(2024)\citenamefont {Abbott} \emph {et~al.}}]{ref:Abbott2024}%
  \BibitemOpen
  \bibfield  {author} {\bibinfo {author} {\bibfnamefont {R.}~\bibnamefont {Abbott}} \emph {et~al.} (\bibinfo {collaboration} {The LIGO Scientific Collaboration and the Virgo Collaboration}),\ }\bibfield  {title} {\bibinfo {title} {{GWTC-2.1}: Deep extended catalog of compact binary coalescences observed by {LIGO} and {Virgo} during the first half of the third observing run},\ }\href {https://doi.org/10.1103/PhysRevD.109.022001} {\bibfield  {journal} {\bibinfo  {journal} {Phys. Rev. D}\ }\textbf {\bibinfo {volume} {109}},\ \bibinfo {pages} {022001} (\bibinfo {year} {2024})}\BibitemShut {NoStop}%
\bibitem [{\citenamefont {Abbott}\ \emph {et~al.}(2023)\citenamefont {Abbott} \emph {et~al.}}]{ref:Abbott2023}%
  \BibitemOpen
  \bibfield  {author} {\bibinfo {author} {\bibfnamefont {R.}~\bibnamefont {Abbott}} \emph {et~al.} (\bibinfo {collaboration} {LIGO Scientific Collaboration, Virgo Collaboration, and KAGRA Collaboration}),\ }\bibfield  {title} {\bibinfo {title} {{GWTC-3}: Compact binary coalescences observed by {LIGO} and {Virgo} during the second part of the third observing run},\ }\href {https://doi.org/10.1103/PhysRevX.13.041039} {\bibfield  {journal} {\bibinfo  {journal} {Phys. Rev. X}\ }\textbf {\bibinfo {volume} {13}},\ \bibinfo {pages} {041039} (\bibinfo {year} {2023})}\BibitemShut {NoStop}%
\bibitem [{\citenamefont {Janka}(2012)}]{ref:Janka2012}%
  \BibitemOpen
  \bibfield  {author} {\bibinfo {author} {\bibfnamefont {H.-T.}\ \bibnamefont {Janka}},\ }\bibfield  {title} {\bibinfo {title} {Explosion mechanisms of core-collapse supernovae},\ }\href {https://doi.org/10.1146/annurev-nucl-102711-094901} {\bibfield  {journal} {\bibinfo  {journal} {Annu. Rev. Nucl. Part. Sci.}\ }\textbf {\bibinfo {volume} {62}},\ \bibinfo {pages} {407} (\bibinfo {year} {2012})}\BibitemShut {NoStop}%
\bibitem [{\citenamefont {Mezzacappa}\ and\ \citenamefont {Zanolin}(2024)}]{ref:Mezzacappa2024}%
  \BibitemOpen
  \bibfield  {author} {\bibinfo {author} {\bibfnamefont {A.}~\bibnamefont {Mezzacappa}}\ and\ \bibinfo {author} {\bibfnamefont {M.}~\bibnamefont {Zanolin}},\ }\href@noop {} {\bibinfo {title} {{Gravitational waves from neutrino-driven core collapse supernovae: Predictions, detection, and parameter estimation}}} (\bibinfo {year} {2024}),\ \Eprint {https://arxiv.org/abs/2401.11635} {arXiv:2401.11635} \BibitemShut {NoStop}%
\bibitem [{\citenamefont {Aasi}\ \emph {et~al.}(2015)\citenamefont {Aasi} \emph {et~al.}}]{ref:Aasi2015}%
  \BibitemOpen
  \bibfield  {author} {\bibinfo {author} {\bibfnamefont {J.}~\bibnamefont {Aasi}} \emph {et~al.} (\bibinfo {collaboration} {LIGO Scientific Collaboration}),\ }\bibfield  {title} {\bibinfo {title} {Advanced {LIGO}},\ }\href {https://doi.org/10.1088/0264-9381/32/7/074001} {\bibfield  {journal} {\bibinfo  {journal} {Classical Quantum Gravity}\ }\textbf {\bibinfo {volume} {32}},\ \bibinfo {pages} {074001} (\bibinfo {year} {2015})}\BibitemShut {NoStop}%
\bibitem [{\citenamefont {Acernese}\ \emph {et~al.}(2014)\citenamefont {Acernese} \emph {et~al.}}]{ref:Acernese2015}%
  \BibitemOpen
  \bibfield  {author} {\bibinfo {author} {\bibfnamefont {F.}~\bibnamefont {Acernese}} \emph {et~al.},\ }\bibfield  {title} {\bibinfo {title} {{Advanced Virgo: A second-generation interferometric gravitational wave detector}},\ }\href {https://doi.org/10.1088/0264-9381/32/2/024001} {\bibfield  {journal} {\bibinfo  {journal} {Classical Quantum Gravity}\ }\textbf {\bibinfo {volume} {32}},\ \bibinfo {pages} {024001} (\bibinfo {year} {2014})}\BibitemShut {NoStop}%
\bibitem [{\citenamefont {Akutsu}\ \emph {et~al.}(2019)\citenamefont {Akutsu} \emph {et~al.}}]{ref:Akutsu2019}%
  \BibitemOpen
  \bibfield  {author} {\bibinfo {author} {\bibfnamefont {T.}~\bibnamefont {Akutsu}} \emph {et~al.},\ }\bibfield  {title} {\bibinfo {title} {{KAGRA}: 2.5 generation interferometric gravitational wave detector},\ }\href {https://doi.org/10.1038/s41550-018-0658-y} {\bibfield  {journal} {\bibinfo  {journal} {Nat. Astron.}\ }\textbf {\bibinfo {volume} {3}},\ \bibinfo {pages} {35} (\bibinfo {year} {2019})}\BibitemShut {NoStop}%
\bibitem [{\citenamefont {Szczepa\ifmmode~\acute{n}\else \'{n}\fi{}czyk}\ \emph {et~al.}(2021)\citenamefont {Szczepa\ifmmode~\acute{n}\else \'{n}\fi{}czyk}, \citenamefont {Antelis}, \citenamefont {Benjamin}, \citenamefont {Cavagli\`a}, \citenamefont {Gondek-Rosi\ifmmode~\acute{n}\else \'{n}\fi{}ska}, \citenamefont {Hansen}, \citenamefont {Klimenko}, \citenamefont {Morales}, \citenamefont {Moreno}, \citenamefont {Mukherjee}, \citenamefont {Nurbek}, \citenamefont {Powell}, \citenamefont {Singh}, \citenamefont {Sitmukhambetov}, \citenamefont {Szewczyk}, \citenamefont {Valdez}, \citenamefont {Vedovato}, \citenamefont {Westhouse}, \citenamefont {Zanolin},\ and\ \citenamefont {Zheng}}]{ref:Marek2021}%
  \BibitemOpen
  \bibfield  {author} {\bibinfo {author} {\bibfnamefont {M.~J.}\ \bibnamefont {Szczepa\ifmmode~\acute{n}\else \'{n}\fi{}czyk}}, \bibinfo {author} {\bibfnamefont {J.~M.}\ \bibnamefont {Antelis}}, \bibinfo {author} {\bibfnamefont {M.}~\bibnamefont {Benjamin}}, \bibinfo {author} {\bibfnamefont {M.}~\bibnamefont {Cavagli\`a}}, \bibinfo {author} {\bibfnamefont {D.}~\bibnamefont {Gondek-Rosi\ifmmode~\acute{n}\else \'{n}\fi{}ska}}, \bibinfo {author} {\bibfnamefont {T.}~\bibnamefont {Hansen}}, \bibinfo {author} {\bibfnamefont {S.}~\bibnamefont {Klimenko}}, \bibinfo {author} {\bibfnamefont {M.~D.}\ \bibnamefont {Morales}}, \bibinfo {author} {\bibfnamefont {C.}~\bibnamefont {Moreno}}, \bibinfo {author} {\bibfnamefont {S.}~\bibnamefont {Mukherjee}}, \bibinfo {author} {\bibfnamefont {G.}~\bibnamefont {Nurbek}}, \bibinfo {author} {\bibfnamefont {J.}~\bibnamefont {Powell}}, \bibinfo {author} {\bibfnamefont {N.}~\bibnamefont {Singh}}, \bibinfo {author} {\bibfnamefont {S.}~\bibnamefont {Sitmukhambetov}}, \bibinfo {author}
  {\bibfnamefont {P.}~\bibnamefont {Szewczyk}}, \bibinfo {author} {\bibfnamefont {O.}~\bibnamefont {Valdez}}, \bibinfo {author} {\bibfnamefont {G.}~\bibnamefont {Vedovato}}, \bibinfo {author} {\bibfnamefont {J.}~\bibnamefont {Westhouse}}, \bibinfo {author} {\bibfnamefont {M.}~\bibnamefont {Zanolin}},\ and\ \bibinfo {author} {\bibfnamefont {Y.}~\bibnamefont {Zheng}},\ }\bibfield  {title} {\bibinfo {title} {Detecting and reconstructing gravitational waves from the next galactic core-collapse supernova in the advanced detector era},\ }\href {https://doi.org/10.1103/PhysRevD.104.102002} {\bibfield  {journal} {\bibinfo  {journal} {Phys. Rev. D}\ }\textbf {\bibinfo {volume} {104}},\ \bibinfo {pages} {102002} (\bibinfo {year} {2021})}\BibitemShut {NoStop}%
\bibitem [{\citenamefont {Hayama}\ \emph {et~al.}(2015)\citenamefont {Hayama}, \citenamefont {Kuroda}, \citenamefont {Kotake},\ and\ \citenamefont {Takiwaki}}]{ref:hayama2015coherent}%
  \BibitemOpen
  \bibfield  {author} {\bibinfo {author} {\bibfnamefont {K.}~\bibnamefont {Hayama}}, \bibinfo {author} {\bibfnamefont {T.}~\bibnamefont {Kuroda}}, \bibinfo {author} {\bibfnamefont {K.}~\bibnamefont {Kotake}},\ and\ \bibinfo {author} {\bibfnamefont {T.}~\bibnamefont {Takiwaki}},\ }\bibfield  {title} {\bibinfo {title} {Coherent network analysis of gravitational waves from three-dimensional core-collapse supernova models},\ }\href {https://doi.org/10.1103/PhysRevD.92.122001} {\bibfield  {journal} {\bibinfo  {journal} {Phys. Rev. D}\ }\textbf {\bibinfo {volume} {92}},\ \bibinfo {pages} {122001} (\bibinfo {year} {2015})}\BibitemShut {NoStop}%
\bibitem [{\citenamefont {Abbott}\ \emph {et~al.}(2017)\citenamefont {Abbott} \emph {et~al.}}]{ref:Abbott2017_ce}%
  \BibitemOpen
  \bibfield  {author} {\bibinfo {author} {\bibfnamefont {B.~P.}\ \bibnamefont {Abbott}} \emph {et~al.},\ }\bibfield  {title} {\bibinfo {title} {Exploring the sensitivity of next generation gravitational wave detectors},\ }\href {https://doi.org/10.1088/1361-6382/aa51f4} {\bibfield  {journal} {\bibinfo  {journal} {Classical Quantum Gravity}\ }\textbf {\bibinfo {volume} {34}},\ \bibinfo {pages} {044001} (\bibinfo {year} {2017})}\BibitemShut {NoStop}%
\bibitem [{\citenamefont {Reitze}\ \emph {et~al.}(2019)\citenamefont {Reitze}, \citenamefont {Adhikari}, \citenamefont {Ballmer}, \citenamefont {Barish}, \citenamefont {Barsotti}, \citenamefont {Billingsley}, \citenamefont {Brown}, \citenamefont {Chen}, \citenamefont {Coyne}, \citenamefont {Eisenstein}, \citenamefont {Evans}, \citenamefont {Fritschel}, \citenamefont {Hall}, \citenamefont {Lazzarini}, \citenamefont {Lovelace}, \citenamefont {Read}, \citenamefont {Sathyaprakash}, \citenamefont {Shoemaker}, \citenamefont {Smith}, \citenamefont {Torrie}, \citenamefont {Vitale}, \citenamefont {Weiss}, \citenamefont {Wipf},\ and\ \citenamefont {Zucker}}]{ref:Reitze2019}%
  \BibitemOpen
  \bibfield  {author} {\bibinfo {author} {\bibfnamefont {D.}~\bibnamefont {Reitze}}, \bibinfo {author} {\bibfnamefont {R.~X.}\ \bibnamefont {Adhikari}}, \bibinfo {author} {\bibfnamefont {S.}~\bibnamefont {Ballmer}}, \bibinfo {author} {\bibfnamefont {B.}~\bibnamefont {Barish}}, \bibinfo {author} {\bibfnamefont {L.}~\bibnamefont {Barsotti}}, \bibinfo {author} {\bibfnamefont {G.}~\bibnamefont {Billingsley}}, \bibinfo {author} {\bibfnamefont {D.~A.}\ \bibnamefont {Brown}}, \bibinfo {author} {\bibfnamefont {Y.}~\bibnamefont {Chen}}, \bibinfo {author} {\bibfnamefont {D.}~\bibnamefont {Coyne}}, \bibinfo {author} {\bibfnamefont {R.}~\bibnamefont {Eisenstein}}, \bibinfo {author} {\bibfnamefont {M.}~\bibnamefont {Evans}}, \bibinfo {author} {\bibfnamefont {P.}~\bibnamefont {Fritschel}}, \bibinfo {author} {\bibfnamefont {E.~D.}\ \bibnamefont {Hall}}, \bibinfo {author} {\bibfnamefont {A.}~\bibnamefont {Lazzarini}}, \bibinfo {author} {\bibfnamefont {G.}~\bibnamefont {Lovelace}}, \bibinfo {author} {\bibfnamefont
  {J.}~\bibnamefont {Read}}, \bibinfo {author} {\bibfnamefont {B.~S.}\ \bibnamefont {Sathyaprakash}}, \bibinfo {author} {\bibfnamefont {D.}~\bibnamefont {Shoemaker}}, \bibinfo {author} {\bibfnamefont {J.}~\bibnamefont {Smith}}, \bibinfo {author} {\bibfnamefont {C.}~\bibnamefont {Torrie}}, \bibinfo {author} {\bibfnamefont {S.}~\bibnamefont {Vitale}}, \bibinfo {author} {\bibfnamefont {R.}~\bibnamefont {Weiss}}, \bibinfo {author} {\bibfnamefont {C.}~\bibnamefont {Wipf}},\ and\ \bibinfo {author} {\bibfnamefont {M.}~\bibnamefont {Zucker}},\ }\href {https://ui.adsabs.harvard.edu/abs/2019BAAS...51g..35R} {\bibinfo {title} {{Cosmic Explorer}: {The U.S.} contribution to gravitational-wave astronomy beyond {LIGO}}} (\bibinfo {year} {2019}),\ \Eprint {https://arxiv.org/abs/1907.04833} {arXiv:1907.04833} \BibitemShut {NoStop}%
\bibitem [{\citenamefont {Punturo}\ \emph {et~al.}(2010)\citenamefont {Punturo} \emph {et~al.}}]{ref:Punturo2010}%
  \BibitemOpen
  \bibfield  {author} {\bibinfo {author} {\bibfnamefont {M.}~\bibnamefont {Punturo}} \emph {et~al.},\ }\bibfield  {title} {\bibinfo {title} {{The Einstein Telescope: A third-generation gravitational wave observatory}},\ }\href {https://doi.org/10.1088/0264-9381/27/19/194002} {\bibfield  {journal} {\bibinfo  {journal} {Classical Quantum Gravity}\ }\textbf {\bibinfo {volume} {27}},\ \bibinfo {pages} {194002} (\bibinfo {year} {2010})}\BibitemShut {NoStop}%
\bibitem [{\citenamefont {Ackley}\ \emph {et~al.}(2020)\citenamefont {Ackley} \emph {et~al.}}]{ref:Ackley2020}%
  \BibitemOpen
  \bibfield  {author} {\bibinfo {author} {\bibfnamefont {K.}~\bibnamefont {Ackley}} \emph {et~al.},\ }\bibfield  {title} {\bibinfo {title} {{Neutron star extreme matter observatory: A kilohertz-band gravitational-wave detector in the global network}},\ }\href {https://doi.org/10.1017/pasa.2020.39} {\bibfield  {journal} {\bibinfo  {journal} {Pub. Astron. Soc. Aust.}\ }\textbf {\bibinfo {volume} {37}},\ \bibinfo {pages} {e047} (\bibinfo {year} {2020})}\BibitemShut {NoStop}%
\bibitem [{\citenamefont {Szczepa{\'n}czyk}\ and\ \citenamefont {Zanolin}(2022)}]{ref:szczepanczyk2022gravitational}%
  \BibitemOpen
  \bibfield  {author} {\bibinfo {author} {\bibfnamefont {M.}~\bibnamefont {Szczepa{\'n}czyk}}\ and\ \bibinfo {author} {\bibfnamefont {M.}~\bibnamefont {Zanolin}},\ }\bibfield  {title} {\bibinfo {title} {Gravitational waves from a core-collapse supernova: Perspectives with detectors in the late 2020s and early 2030s},\ }\href {https://doi.org/10.3390/galaxies10030070} {\bibfield  {journal} {\bibinfo  {journal} {Galaxies}\ }\textbf {\bibinfo {volume} {10}},\ \bibinfo {pages} {70} (\bibinfo {year} {2022})}\BibitemShut {NoStop}%
\bibitem [{\citenamefont {Klimenko}\ \emph {et~al.}(2008)\citenamefont {Klimenko}, \citenamefont {Yakushin}, \citenamefont {Mercer},\ and\ \citenamefont {Mitselmakher}}]{ref:Klimenko2008}%
  \BibitemOpen
  \bibfield  {author} {\bibinfo {author} {\bibfnamefont {S.}~\bibnamefont {Klimenko}}, \bibinfo {author} {\bibfnamefont {I.}~\bibnamefont {Yakushin}}, \bibinfo {author} {\bibfnamefont {A.}~\bibnamefont {Mercer}},\ and\ \bibinfo {author} {\bibfnamefont {G.}~\bibnamefont {Mitselmakher}},\ }\bibfield  {title} {\bibinfo {title} {A coherent method for detection of gravitational wave bursts},\ }\href {https://doi.org/10.1088/0264-9381/25/11/114029} {\bibfield  {journal} {\bibinfo  {journal} {Classical Quantum Gravity}\ }\textbf {\bibinfo {volume} {25}},\ \bibinfo {pages} {114029} (\bibinfo {year} {2008})}\BibitemShut {NoStop}%
\bibitem [{\citenamefont {Klimenko}\ \emph {et~al.}(2016)\citenamefont {Klimenko}, \citenamefont {Vedovato}, \citenamefont {Drago}, \citenamefont {Salemi}, \citenamefont {Tiwari}, \citenamefont {Prodi}, \citenamefont {Lazzaro}, \citenamefont {Ackley}, \citenamefont {Tiwari}, \citenamefont {Da~Silva},\ and\ \citenamefont {Mitselmakher}}]{ref:Klimenko2016}%
  \BibitemOpen
  \bibfield  {author} {\bibinfo {author} {\bibfnamefont {S.}~\bibnamefont {Klimenko}}, \bibinfo {author} {\bibfnamefont {G.}~\bibnamefont {Vedovato}}, \bibinfo {author} {\bibfnamefont {M.}~\bibnamefont {Drago}}, \bibinfo {author} {\bibfnamefont {F.}~\bibnamefont {Salemi}}, \bibinfo {author} {\bibfnamefont {V.}~\bibnamefont {Tiwari}}, \bibinfo {author} {\bibfnamefont {G.~A.}\ \bibnamefont {Prodi}}, \bibinfo {author} {\bibfnamefont {C.}~\bibnamefont {Lazzaro}}, \bibinfo {author} {\bibfnamefont {K.}~\bibnamefont {Ackley}}, \bibinfo {author} {\bibfnamefont {S.}~\bibnamefont {Tiwari}}, \bibinfo {author} {\bibfnamefont {C.~F.}\ \bibnamefont {Da~Silva}},\ and\ \bibinfo {author} {\bibfnamefont {G.}~\bibnamefont {Mitselmakher}},\ }\bibfield  {title} {\bibinfo {title} {{Method for detection and reconstruction of gravitational wave transients with networks of advanced detectors}},\ }\href {https://doi.org/10.1103/PhysRevD.93.042004} {\bibfield  {journal} {\bibinfo  {journal} {Phys. Rev. D}\ }\textbf {\bibinfo {volume}
  {93}},\ \bibinfo {pages} {042004} (\bibinfo {year} {2016})}\BibitemShut {NoStop}%
\bibitem [{\citenamefont {Drago}\ \emph {et~al.}(2021)\citenamefont {Drago} \emph {et~al.}}]{ref:Drago2021}%
  \BibitemOpen
  \bibfield  {author} {\bibinfo {author} {\bibfnamefont {M.}~\bibnamefont {Drago}} \emph {et~al.},\ }\bibfield  {title} {\bibinfo {title} {{Coherent WaveBurst, A pipeline for unmodeled gravitational-wave data analysis}},\ }\href {https://doi.org/https://doi.org/10.1016/j.softx.2021.100678} {\bibfield  {journal} {\bibinfo  {journal} {SoftwareX}\ }\textbf {\bibinfo {volume} {14}},\ \bibinfo {pages} {100678} (\bibinfo {year} {2021})}\BibitemShut {NoStop}%
\bibitem [{\citenamefont {Astone}\ \emph {et~al.}(2018)\citenamefont {Astone}, \citenamefont {Cerd\'a-Dur\'an}, \citenamefont {Di~Palma}, \citenamefont {Drago}, \citenamefont {Muciaccia}, \citenamefont {Palomba},\ and\ \citenamefont {Ricci}}]{ref:Astone2018}%
  \BibitemOpen
  \bibfield  {author} {\bibinfo {author} {\bibfnamefont {P.}~\bibnamefont {Astone}}, \bibinfo {author} {\bibfnamefont {P.}~\bibnamefont {Cerd\'a-Dur\'an}}, \bibinfo {author} {\bibfnamefont {I.}~\bibnamefont {Di~Palma}}, \bibinfo {author} {\bibfnamefont {M.}~\bibnamefont {Drago}}, \bibinfo {author} {\bibfnamefont {F.}~\bibnamefont {Muciaccia}}, \bibinfo {author} {\bibfnamefont {C.}~\bibnamefont {Palomba}},\ and\ \bibinfo {author} {\bibfnamefont {F.}~\bibnamefont {Ricci}},\ }\bibfield  {title} {\bibinfo {title} {New method to observe gravitational waves emitted by core collapse supernovae},\ }\href {https://doi.org/10.1103/PhysRevD.98.122002} {\bibfield  {journal} {\bibinfo  {journal} {Phys. Rev. D}\ }\textbf {\bibinfo {volume} {98}},\ \bibinfo {pages} {122002} (\bibinfo {year} {2018})}\BibitemShut {NoStop}%
\bibitem [{\citenamefont {Iess}\ \emph {et~al.}(2020)\citenamefont {Iess}, \citenamefont {Cuoco}, \citenamefont {Morawski},\ and\ \citenamefont {Powell}}]{ref:Iess2020}%
  \BibitemOpen
  \bibfield  {author} {\bibinfo {author} {\bibfnamefont {A.}~\bibnamefont {Iess}}, \bibinfo {author} {\bibfnamefont {E.}~\bibnamefont {Cuoco}}, \bibinfo {author} {\bibfnamefont {F.}~\bibnamefont {Morawski}},\ and\ \bibinfo {author} {\bibfnamefont {J.}~\bibnamefont {Powell}},\ }\bibfield  {title} {\bibinfo {title} {Core-collapse supernova gravitational-wave search and deep learning classification},\ }\href {https://doi.org/10.1088/2632-2153/ab7d31} {\bibfield  {journal} {\bibinfo  {journal} {Mach. Learn.}\ }\textbf {\bibinfo {volume} {1}},\ \bibinfo {pages} {025014} (\bibinfo {year} {2020})}\BibitemShut {NoStop}%
\bibitem [{\citenamefont {Chan}\ \emph {et~al.}(2020)\citenamefont {Chan}, \citenamefont {Heng},\ and\ \citenamefont {Messenger}}]{ref:Chan2020}%
  \BibitemOpen
  \bibfield  {author} {\bibinfo {author} {\bibfnamefont {M.~L.}\ \bibnamefont {Chan}}, \bibinfo {author} {\bibfnamefont {I.~S.}\ \bibnamefont {Heng}},\ and\ \bibinfo {author} {\bibfnamefont {C.}~\bibnamefont {Messenger}},\ }\bibfield  {title} {\bibinfo {title} {Detection and classification of supernova gravitational wave signals: A deep learning approach},\ }\href {https://doi.org/10.1103/PhysRevD.102.043022} {\bibfield  {journal} {\bibinfo  {journal} {Phys. Rev. D}\ }\textbf {\bibinfo {volume} {102}},\ \bibinfo {pages} {043022} (\bibinfo {year} {2020})}\BibitemShut {NoStop}%
\bibitem [{\citenamefont {L\'opez}\ \emph {et~al.}(2021)\citenamefont {L\'opez}, \citenamefont {Di~Palma}, \citenamefont {Drago}, \citenamefont {Cerd\'a-Dur\'an},\ and\ \citenamefont {Ricci}}]{ref:Lopez2021}%
  \BibitemOpen
  \bibfield  {author} {\bibinfo {author} {\bibfnamefont {M.}~\bibnamefont {L\'opez}}, \bibinfo {author} {\bibfnamefont {I.}~\bibnamefont {Di~Palma}}, \bibinfo {author} {\bibfnamefont {M.}~\bibnamefont {Drago}}, \bibinfo {author} {\bibfnamefont {P.}~\bibnamefont {Cerd\'a-Dur\'an}},\ and\ \bibinfo {author} {\bibfnamefont {F.}~\bibnamefont {Ricci}},\ }\bibfield  {title} {\bibinfo {title} {Deep learning for core-collapse supernova detection},\ }\href {https://doi.org/10.1103/PhysRevD.103.063011} {\bibfield  {journal} {\bibinfo  {journal} {Phys. Rev. D}\ }\textbf {\bibinfo {volume} {103}},\ \bibinfo {pages} {063011} (\bibinfo {year} {2021})}\BibitemShut {NoStop}%
\bibitem [{\citenamefont {Iess}\ \emph {et~al.}(2023)\citenamefont {Iess}, \citenamefont {Cuoco}, \citenamefont {Morawski}, \citenamefont {Nicolaou},\ and\ \citenamefont {Lahav}}]{ref:Iess2023}%
  \BibitemOpen
  \bibfield  {author} {\bibinfo {author} {\bibfnamefont {A.}~\bibnamefont {Iess}}, \bibinfo {author} {\bibfnamefont {E.}~\bibnamefont {Cuoco}}, \bibinfo {author} {\bibfnamefont {F.}~\bibnamefont {Morawski}}, \bibinfo {author} {\bibfnamefont {C.}~\bibnamefont {Nicolaou}},\ and\ \bibinfo {author} {\bibfnamefont {O.}~\bibnamefont {Lahav}},\ }\bibfield  {title} {\bibinfo {title} {{LSTM} and {CNN} application for core-collapse supernova search in gravitational wave real data},\ }\href {https://doi.org/10.1051/0004-6361/202142525} {\bibfield  {journal} {\bibinfo  {journal} {Astron. Astrophys.}\ }\textbf {\bibinfo {volume} {669}},\ \bibinfo {pages} {A42} (\bibinfo {year} {2023})}\BibitemShut {NoStop}%
\bibitem [{\citenamefont {Sasaoka}\ \emph {et~al.}(2023)\citenamefont {Sasaoka}, \citenamefont {Koyama}, \citenamefont {Dominguez}, \citenamefont {Sakai}, \citenamefont {Somiya}, \citenamefont {Omae},\ and\ \citenamefont {Takahashi}}]{ref:Sasaoka2023}%
  \BibitemOpen
  \bibfield  {author} {\bibinfo {author} {\bibfnamefont {S.}~\bibnamefont {Sasaoka}}, \bibinfo {author} {\bibfnamefont {N.}~\bibnamefont {Koyama}}, \bibinfo {author} {\bibfnamefont {D.}~\bibnamefont {Dominguez}}, \bibinfo {author} {\bibfnamefont {Y.}~\bibnamefont {Sakai}}, \bibinfo {author} {\bibfnamefont {K.}~\bibnamefont {Somiya}}, \bibinfo {author} {\bibfnamefont {Y.}~\bibnamefont {Omae}},\ and\ \bibinfo {author} {\bibfnamefont {H.}~\bibnamefont {Takahashi}},\ }\bibfield  {title} {\bibinfo {title} {Visualizing convolutional neural network for classifying gravitational waves from core-collapse supernovae},\ }\href {https://doi.org/10.1103/PhysRevD.108.123033} {\bibfield  {journal} {\bibinfo  {journal} {Phys. Rev. D}\ }\textbf {\bibinfo {volume} {108}},\ \bibinfo {pages} {123033} (\bibinfo {year} {2023})}\BibitemShut {NoStop}%
\bibitem [{\citenamefont {Edwards}(2021)}]{ref:Edwards2021}%
  \BibitemOpen
  \bibfield  {author} {\bibinfo {author} {\bibfnamefont {M.~C.}\ \bibnamefont {Edwards}},\ }\bibfield  {title} {\bibinfo {title} {Classifying the equation of state from rotating core collapse gravitational waves with deep learning},\ }\href {https://doi.org/10.1103/PhysRevD.103.024025} {\bibfield  {journal} {\bibinfo  {journal} {Phys. Rev. D}\ }\textbf {\bibinfo {volume} {103}},\ \bibinfo {pages} {024025} (\bibinfo {year} {2021})}\BibitemShut {NoStop}%
\bibitem [{\citenamefont {Saiz-Pérez}\ \emph {et~al.}(2022)\citenamefont {Saiz-Pérez}, \citenamefont {Torres-Forné},\ and\ \citenamefont {Font}}]{ref:Perez2022}%
  \BibitemOpen
  \bibfield  {author} {\bibinfo {author} {\bibfnamefont {A.}~\bibnamefont {Saiz-Pérez}}, \bibinfo {author} {\bibfnamefont {A.}~\bibnamefont {Torres-Forné}},\ and\ \bibinfo {author} {\bibfnamefont {J.~A.}\ \bibnamefont {Font}},\ }\bibfield  {title} {\bibinfo {title} {Classification of core-collapse supernova explosions with learned dictionaries},\ }\href {https://doi.org/10.1093/mnras/stac698} {\bibfield  {journal} {\bibinfo  {journal} {Mon. Not. R. Astron. Soc.}\ }\textbf {\bibinfo {volume} {512}},\ \bibinfo {pages} {3815} (\bibinfo {year} {2022})}\BibitemShut {NoStop}%
\bibitem [{\citenamefont {Mitra}\ \emph {et~al.}(2024)\citenamefont {Mitra}, \citenamefont {Orel}, \citenamefont {Abylkairov}, \citenamefont {Shukirgaliyev},\ and\ \citenamefont {Abdikamalov}}]{ref:Mitra2024}%
  \BibitemOpen
  \bibfield  {author} {\bibinfo {author} {\bibfnamefont {A.}~\bibnamefont {Mitra}}, \bibinfo {author} {\bibfnamefont {D.}~\bibnamefont {Orel}}, \bibinfo {author} {\bibfnamefont {Y.~S.}\ \bibnamefont {Abylkairov}}, \bibinfo {author} {\bibfnamefont {B.}~\bibnamefont {Shukirgaliyev}},\ and\ \bibinfo {author} {\bibfnamefont {E.}~\bibnamefont {Abdikamalov}},\ }\bibfield  {title} {\bibinfo {title} {Probing nuclear physics with supernova gravitational waves and machine learning},\ }\href {https://doi.org/10.1093/mnras/stae714} {\bibfield  {journal} {\bibinfo  {journal} {Mon. Not. R. Astron. Soc.}\ }\textbf {\bibinfo {volume} {529}},\ \bibinfo {pages} {3582} (\bibinfo {year} {2024})}\BibitemShut {NoStop}%
\bibitem [{\citenamefont {Powell}\ \emph {et~al.}(2024)\citenamefont {Powell}, \citenamefont {Iess}, \citenamefont {Llorens-Monteagudo}, \citenamefont {Obergaulinger}, \citenamefont {M\"uller}, \citenamefont {Torres-Forn\'e}, \citenamefont {Cuoco},\ and\ \citenamefont {Font}}]{ref:Powell2024}%
  \BibitemOpen
  \bibfield  {author} {\bibinfo {author} {\bibfnamefont {J.}~\bibnamefont {Powell}}, \bibinfo {author} {\bibfnamefont {A.}~\bibnamefont {Iess}}, \bibinfo {author} {\bibfnamefont {M.}~\bibnamefont {Llorens-Monteagudo}}, \bibinfo {author} {\bibfnamefont {M.}~\bibnamefont {Obergaulinger}}, \bibinfo {author} {\bibfnamefont {B.}~\bibnamefont {M\"uller}}, \bibinfo {author} {\bibfnamefont {A.}~\bibnamefont {Torres-Forn\'e}}, \bibinfo {author} {\bibfnamefont {E.}~\bibnamefont {Cuoco}},\ and\ \bibinfo {author} {\bibfnamefont {J.~A.}\ \bibnamefont {Font}},\ }\bibfield  {title} {\bibinfo {title} {Determining the core-collapse supernova explosion mechanism with current and future gravitational-wave observatories},\ }\href {https://doi.org/10.1103/PhysRevD.109.063019} {\bibfield  {journal} {\bibinfo  {journal} {Phys. Rev. D}\ }\textbf {\bibinfo {volume} {109}},\ \bibinfo {pages} {063019} (\bibinfo {year} {2024})}\BibitemShut {NoStop}%
\bibitem [{\citenamefont {Andersson}\ and\ \citenamefont {Kokkotas}(1998)}]{ref:Andersson1998}%
  \BibitemOpen
  \bibfield  {author} {\bibinfo {author} {\bibfnamefont {N.}~\bibnamefont {Andersson}}\ and\ \bibinfo {author} {\bibfnamefont {K.~D.}\ \bibnamefont {Kokkotas}},\ }\bibfield  {title} {\bibinfo {title} {{Towards gravitational wave asteroseismology}},\ }\href {https://doi.org/10.1046/j.1365-8711.1998.01840.x} {\bibfield  {journal} {\bibinfo  {journal} {Mon. Not. R. Astron. Soc.}\ }\textbf {\bibinfo {volume} {299}},\ \bibinfo {pages} {1059} (\bibinfo {year} {1998})}\BibitemShut {NoStop}%
\bibitem [{\citenamefont {Sotani}\ and\ \citenamefont {Takiwaki}(2016)}]{ref:Sotani2016}%
  \BibitemOpen
  \bibfield  {author} {\bibinfo {author} {\bibfnamefont {H.}~\bibnamefont {Sotani}}\ and\ \bibinfo {author} {\bibfnamefont {T.}~\bibnamefont {Takiwaki}},\ }\bibfield  {title} {\bibinfo {title} {Gravitational wave asteroseismology with protoneutron stars},\ }\href {https://doi.org/10.1103/PhysRevD.94.044043} {\bibfield  {journal} {\bibinfo  {journal} {Phys. Rev. D}\ }\textbf {\bibinfo {volume} {94}},\ \bibinfo {pages} {044043} (\bibinfo {year} {2016})}\BibitemShut {NoStop}%
\bibitem [{\citenamefont {Sotani}\ \emph {et~al.}(2019)\citenamefont {Sotani}, \citenamefont {Kuroda}, \citenamefont {Takiwaki},\ and\ \citenamefont {Kotake}}]{ref:Sotani2019}%
  \BibitemOpen
  \bibfield  {author} {\bibinfo {author} {\bibfnamefont {H.}~\bibnamefont {Sotani}}, \bibinfo {author} {\bibfnamefont {T.}~\bibnamefont {Kuroda}}, \bibinfo {author} {\bibfnamefont {T.}~\bibnamefont {Takiwaki}},\ and\ \bibinfo {author} {\bibfnamefont {K.}~\bibnamefont {Kotake}},\ }\bibfield  {title} {\bibinfo {title} {Dependence of the outer boundary condition on protoneutron star asteroseismology with gravitational-wave signatures},\ }\href {https://doi.org/10.1103/PhysRevD.99.123024} {\bibfield  {journal} {\bibinfo  {journal} {Phys. Rev. D}\ }\textbf {\bibinfo {volume} {99}},\ \bibinfo {pages} {123024} (\bibinfo {year} {2019})}\BibitemShut {NoStop}%
\bibitem [{\citenamefont {Torres-Forn\'e}\ \emph {et~al.}(2019)\citenamefont {Torres-Forn\'e}, \citenamefont {Cerd\'a-Dur\'an}, \citenamefont {Obergaulinger}, \citenamefont {M\"uller},\ and\ \citenamefont {Font}}]{ref:Torres2019}%
  \BibitemOpen
  \bibfield  {author} {\bibinfo {author} {\bibfnamefont {A.}~\bibnamefont {Torres-Forn\'e}}, \bibinfo {author} {\bibfnamefont {P.}~\bibnamefont {Cerd\'a-Dur\'an}}, \bibinfo {author} {\bibfnamefont {M.}~\bibnamefont {Obergaulinger}}, \bibinfo {author} {\bibfnamefont {B.}~\bibnamefont {M\"uller}},\ and\ \bibinfo {author} {\bibfnamefont {J.~A.}\ \bibnamefont {Font}},\ }\bibfield  {title} {\bibinfo {title} {Universal relations for gravitational-wave asteroseismology of protoneutron stars},\ }\href {https://doi.org/10.1103/PhysRevLett.123.051102} {\bibfield  {journal} {\bibinfo  {journal} {Phys. Rev. Lett.}\ }\textbf {\bibinfo {volume} {123}},\ \bibinfo {pages} {051102} (\bibinfo {year} {2019})}\BibitemShut {NoStop}%
\bibitem [{\citenamefont {Torres-Forn\'e}\ \emph {et~al.}(2021)\citenamefont {Torres-Forn\'e}, \citenamefont {Cerd\'a-Dur\'an}, \citenamefont {Obergaulinger}, \citenamefont {M\"uller},\ and\ \citenamefont {Font}}]{ref:Torres2021}%
  \BibitemOpen
  \bibfield  {author} {\bibinfo {author} {\bibfnamefont {A.}~\bibnamefont {Torres-Forn\'e}}, \bibinfo {author} {\bibfnamefont {P.}~\bibnamefont {Cerd\'a-Dur\'an}}, \bibinfo {author} {\bibfnamefont {M.}~\bibnamefont {Obergaulinger}}, \bibinfo {author} {\bibfnamefont {B.}~\bibnamefont {M\"uller}},\ and\ \bibinfo {author} {\bibfnamefont {J.~A.}\ \bibnamefont {Font}},\ }\bibfield  {title} {\bibinfo {title} {{Erratum: Universal relations for gravitational-wave asteroseismology of protoneutron stars [Phys. Rev. Lett. 123, 051102 (2019)]}},\ }\href {https://doi.org/10.1103/PhysRevLett.127.239901} {\bibfield  {journal} {\bibinfo  {journal} {Phys. Rev. Lett.}\ }\textbf {\bibinfo {volume} {127}},\ \bibinfo {pages} {239901} (\bibinfo {year} {2021})}\BibitemShut {NoStop}%
\bibitem [{\citenamefont {Sotani}\ \emph {et~al.}(2021)\citenamefont {Sotani}, \citenamefont {Takiwaki},\ and\ \citenamefont {Togashi}}]{ref:Sotani2021}%
  \BibitemOpen
  \bibfield  {author} {\bibinfo {author} {\bibfnamefont {H.}~\bibnamefont {Sotani}}, \bibinfo {author} {\bibfnamefont {T.}~\bibnamefont {Takiwaki}},\ and\ \bibinfo {author} {\bibfnamefont {H.}~\bibnamefont {Togashi}},\ }\bibfield  {title} {\bibinfo {title} {Universal relation for supernova gravitational waves},\ }\href {https://doi.org/10.1103/PhysRevD.104.123009} {\bibfield  {journal} {\bibinfo  {journal} {Phys. Rev. D}\ }\textbf {\bibinfo {volume} {104}},\ \bibinfo {pages} {123009} (\bibinfo {year} {2021})}\BibitemShut {NoStop}%
\bibitem [{\citenamefont {Bizouard}\ \emph {et~al.}(2021{\natexlab{a}})\citenamefont {Bizouard}, \citenamefont {Maturana-Russel}, \citenamefont {Torres-Forn\'e}, \citenamefont {Obergaulinger}, \citenamefont {Cerd\'a-Dur\'an}, \citenamefont {Christensen}, \citenamefont {Font},\ and\ \citenamefont {Meyer}}]{ref:Bizouard2021}%
  \BibitemOpen
  \bibfield  {author} {\bibinfo {author} {\bibfnamefont {M.-A.}\ \bibnamefont {Bizouard}}, \bibinfo {author} {\bibfnamefont {P.}~\bibnamefont {Maturana-Russel}}, \bibinfo {author} {\bibfnamefont {A.}~\bibnamefont {Torres-Forn\'e}}, \bibinfo {author} {\bibfnamefont {M.}~\bibnamefont {Obergaulinger}}, \bibinfo {author} {\bibfnamefont {P.}~\bibnamefont {Cerd\'a-Dur\'an}}, \bibinfo {author} {\bibfnamefont {N.}~\bibnamefont {Christensen}}, \bibinfo {author} {\bibfnamefont {J.~A.}\ \bibnamefont {Font}},\ and\ \bibinfo {author} {\bibfnamefont {R.}~\bibnamefont {Meyer}},\ }\bibfield  {title} {\bibinfo {title} {Inference of protoneutron star properties from gravitational-wave data in core-collapse supernovae},\ }\href {https://doi.org/10.1103/PhysRevD.103.063006} {\bibfield  {journal} {\bibinfo  {journal} {Phys. Rev. D}\ }\textbf {\bibinfo {volume} {103}},\ \bibinfo {pages} {063006} (\bibinfo {year} {2021}{\natexlab{a}})}\BibitemShut {NoStop}%
\bibitem [{\citenamefont {Powell}\ and\ \citenamefont {M\"uller}(2022)}]{ref:Powell2022}%
  \BibitemOpen
  \bibfield  {author} {\bibinfo {author} {\bibfnamefont {J.}~\bibnamefont {Powell}}\ and\ \bibinfo {author} {\bibfnamefont {B.}~\bibnamefont {M\"uller}},\ }\bibfield  {title} {\bibinfo {title} {Inferring astrophysical parameters of core-collapse supernovae from their gravitational-wave emission},\ }\href {https://doi.org/10.1103/PhysRevD.105.063018} {\bibfield  {journal} {\bibinfo  {journal} {Phys. Rev. D}\ }\textbf {\bibinfo {volume} {105}},\ \bibinfo {pages} {063018} (\bibinfo {year} {2022})}\BibitemShut {NoStop}%
\bibitem [{\citenamefont {Bruel}\ \emph {et~al.}(2023)\citenamefont {Bruel}, \citenamefont {Bizouard}, \citenamefont {Obergaulinger}, \citenamefont {Maturana-Russel}, \citenamefont {Torres-Forn\'e}, \citenamefont {Cerd\'a-Dur\'an}, \citenamefont {Christensen}, \citenamefont {Font},\ and\ \citenamefont {Meyer}}]{ref:Bruel2023}%
  \BibitemOpen
  \bibfield  {author} {\bibinfo {author} {\bibfnamefont {T.}~\bibnamefont {Bruel}}, \bibinfo {author} {\bibfnamefont {M.-A.}\ \bibnamefont {Bizouard}}, \bibinfo {author} {\bibfnamefont {M.}~\bibnamefont {Obergaulinger}}, \bibinfo {author} {\bibfnamefont {P.}~\bibnamefont {Maturana-Russel}}, \bibinfo {author} {\bibfnamefont {A.}~\bibnamefont {Torres-Forn\'e}}, \bibinfo {author} {\bibfnamefont {P.}~\bibnamefont {Cerd\'a-Dur\'an}}, \bibinfo {author} {\bibfnamefont {N.}~\bibnamefont {Christensen}}, \bibinfo {author} {\bibfnamefont {J.~A.}\ \bibnamefont {Font}},\ and\ \bibinfo {author} {\bibfnamefont {R.}~\bibnamefont {Meyer}},\ }\bibfield  {title} {\bibinfo {title} {Inference of protoneutron star properties in core-collapse supernovae from a gravitational-wave detector network},\ }\href {https://doi.org/10.1103/PhysRevD.107.083029} {\bibfield  {journal} {\bibinfo  {journal} {Phys. Rev. D}\ }\textbf {\bibinfo {volume} {107}},\ \bibinfo {pages} {083029} (\bibinfo {year} {2023})}\BibitemShut {NoStop}%
\bibitem [{\citenamefont {Casallas-Lagos}\ \emph {et~al.}(2023)\citenamefont {Casallas-Lagos}, \citenamefont {Antelis}, \citenamefont {Moreno}, \citenamefont {Zanolin}, \citenamefont {Mezzacappa},\ and\ \citenamefont {Szczepa\ifmmode~\acute{n}\else \'{n}\fi{}czyk}}]{ref:Lagos2023}%
  \BibitemOpen
  \bibfield  {author} {\bibinfo {author} {\bibfnamefont {A.}~\bibnamefont {Casallas-Lagos}}, \bibinfo {author} {\bibfnamefont {J.~M.}\ \bibnamefont {Antelis}}, \bibinfo {author} {\bibfnamefont {C.}~\bibnamefont {Moreno}}, \bibinfo {author} {\bibfnamefont {M.}~\bibnamefont {Zanolin}}, \bibinfo {author} {\bibfnamefont {A.}~\bibnamefont {Mezzacappa}},\ and\ \bibinfo {author} {\bibfnamefont {M.~J.}\ \bibnamefont {Szczepa\ifmmode~\acute{n}\else \'{n}\fi{}czyk}},\ }\bibfield  {title} {\bibinfo {title} {Characterizing the temporal evolution of the high-frequency gravitational wave emission for a core collapse supernova with laser interferometric data: {A} neural network approach},\ }\href {https://doi.org/10.1103/PhysRevD.108.084027} {\bibfield  {journal} {\bibinfo  {journal} {Phys. Rev. D}\ }\textbf {\bibinfo {volume} {108}},\ \bibinfo {pages} {084027} (\bibinfo {year} {2023})}\BibitemShut {NoStop}%
\bibitem [{\citenamefont {{Huang}}\ \emph {et~al.}(1998)\citenamefont {{Huang}}, \citenamefont {{Shen}}, \citenamefont {{Long}}, \citenamefont {{Wu}}, \citenamefont {{Shih}}, \citenamefont {{Zheng}}, \citenamefont {{Yen}}, \citenamefont {{Tung}},\ and\ \citenamefont {{Liu}}}]{ref:Huang1998}%
  \BibitemOpen
  \bibfield  {author} {\bibinfo {author} {\bibfnamefont {N.~E.}\ \bibnamefont {{Huang}}}, \bibinfo {author} {\bibfnamefont {Z.}~\bibnamefont {{Shen}}}, \bibinfo {author} {\bibfnamefont {S.~R.}\ \bibnamefont {{Long}}}, \bibinfo {author} {\bibfnamefont {M.~C.}\ \bibnamefont {{Wu}}}, \bibinfo {author} {\bibfnamefont {H.~H.}\ \bibnamefont {{Shih}}}, \bibinfo {author} {\bibfnamefont {Q.}~\bibnamefont {{Zheng}}}, \bibinfo {author} {\bibfnamefont {N.~C.}\ \bibnamefont {{Yen}}}, \bibinfo {author} {\bibfnamefont {C.~C.}\ \bibnamefont {{Tung}}},\ and\ \bibinfo {author} {\bibfnamefont {H.~H.}\ \bibnamefont {{Liu}}},\ }\bibfield  {title} {\bibinfo {title} {{The empirical mode decomposition and the Hilbert spectrum for nonlinear and non-stationary time series analysis}},\ }\href {https://doi.org/10.1098/rspa.1998.0193} {\bibfield  {journal} {\bibinfo  {journal} {Proc. R. Soc. A}\ }\textbf {\bibinfo {volume} {454}},\ \bibinfo {pages} {903} (\bibinfo {year} {1998})}\BibitemShut {NoStop}%
\bibitem [{\citenamefont {Camp}\ \emph {et~al.}(2007)\citenamefont {Camp}, \citenamefont {Cannizzo},\ and\ \citenamefont {Numata}}]{ref:Camp2007}%
  \BibitemOpen
  \bibfield  {author} {\bibinfo {author} {\bibfnamefont {J.~B.}\ \bibnamefont {Camp}}, \bibinfo {author} {\bibfnamefont {J.~K.}\ \bibnamefont {Cannizzo}},\ and\ \bibinfo {author} {\bibfnamefont {K.}~\bibnamefont {Numata}},\ }\bibfield  {title} {\bibinfo {title} {{Application of the Hilbert-Huang transform to the search for gravitational waves}},\ }\href {https://doi.org/10.1103/PhysRevD.75.061101} {\bibfield  {journal} {\bibinfo  {journal} {Phys. Rev. D}\ }\textbf {\bibinfo {volume} {75}},\ \bibinfo {pages} {061101} (\bibinfo {year} {2007})}\BibitemShut {NoStop}%
\bibitem [{\citenamefont {Kaneyama}\ \emph {et~al.}(2016)\citenamefont {Kaneyama}, \citenamefont {Oohara}, \citenamefont {Takahashi}, \citenamefont {Sekiguchi}, \citenamefont {Tagoshi},\ and\ \citenamefont {Shibata}}]{ref:Kaneyama2016}%
  \BibitemOpen
  \bibfield  {author} {\bibinfo {author} {\bibfnamefont {M.}~\bibnamefont {Kaneyama}}, \bibinfo {author} {\bibfnamefont {K.}~\bibnamefont {Oohara}}, \bibinfo {author} {\bibfnamefont {H.}~\bibnamefont {Takahashi}}, \bibinfo {author} {\bibfnamefont {Y.}~\bibnamefont {Sekiguchi}}, \bibinfo {author} {\bibfnamefont {H.}~\bibnamefont {Tagoshi}},\ and\ \bibinfo {author} {\bibfnamefont {M.}~\bibnamefont {Shibata}},\ }\bibfield  {title} {\bibinfo {title} {Analysis of gravitational waves from binary neutron star merger by {Hilbert-Huang} transform},\ }\href {https://doi.org/10.1103/PhysRevD.93.123010} {\bibfield  {journal} {\bibinfo  {journal} {Phys. Rev. D}\ }\textbf {\bibinfo {volume} {93}},\ \bibinfo {pages} {123010} (\bibinfo {year} {2016})}\BibitemShut {NoStop}%
\bibitem [{\citenamefont {Yoda}\ \emph {et~al.}(2023)\citenamefont {Yoda}, \citenamefont {Oohara}, \citenamefont {Takahashi},\ and\ \citenamefont {Sakai}}]{ref:Yoda2023}%
  \BibitemOpen
  \bibfield  {author} {\bibinfo {author} {\bibfnamefont {I.}~\bibnamefont {Yoda}}, \bibinfo {author} {\bibfnamefont {K.}~\bibnamefont {Oohara}}, \bibinfo {author} {\bibfnamefont {H.}~\bibnamefont {Takahashi}},\ and\ \bibinfo {author} {\bibfnamefont {K.}~\bibnamefont {Sakai}},\ }\bibfield  {title} {\bibinfo {title} {Precise analysis of gravitational waves from binary neutron star coalescence using {Hilbert-Huang} transform based on {Akima} spline interpolation},\ }\href {https://doi.org/10.1093/ptep/ptad101} {\bibfield  {journal} {\bibinfo  {journal} {Prog. Theor. Exp. Phys.}\ }\textbf {\bibinfo {volume} {2023}},\ \bibinfo {pages} {083E01} (\bibinfo {year} {2023})}\BibitemShut {NoStop}%
\bibitem [{\citenamefont {Sakai}\ \emph {et~al.}(2017)\citenamefont {Sakai}, \citenamefont {Oohara}, \citenamefont {Nakano}, \citenamefont {Kaneyama},\ and\ \citenamefont {Takahashi}}]{ref:Sakai2017}%
  \BibitemOpen
  \bibfield  {author} {\bibinfo {author} {\bibfnamefont {K.}~\bibnamefont {Sakai}}, \bibinfo {author} {\bibfnamefont {K.}~\bibnamefont {Oohara}}, \bibinfo {author} {\bibfnamefont {H.}~\bibnamefont {Nakano}}, \bibinfo {author} {\bibfnamefont {M.}~\bibnamefont {Kaneyama}},\ and\ \bibinfo {author} {\bibfnamefont {H.}~\bibnamefont {Takahashi}},\ }\bibfield  {title} {\bibinfo {title} {{Estimation of starting times of quasinormal modes in ringdown gravitational waves with the Hilbert-Huang transform}},\ }\href {https://doi.org/10.1103/PhysRevD.96.044047} {\bibfield  {journal} {\bibinfo  {journal} {Phys. Rev. D}\ }\textbf {\bibinfo {volume} {96}},\ \bibinfo {pages} {044047} (\bibinfo {year} {2017})}\BibitemShut {NoStop}%
\bibitem [{\citenamefont {Takeda}\ \emph {et~al.}(2021)\citenamefont {Takeda}, \citenamefont {Hiranuma}, \citenamefont {Kanda}, \citenamefont {Kotake}, \citenamefont {Kuroda}, \citenamefont {Negishi}, \citenamefont {Oohara}, \citenamefont {Sakai}, \citenamefont {Sakai}, \citenamefont {Sawada}, \citenamefont {Takahashi}, \citenamefont {Tsuchida}, \citenamefont {Watanabe},\ and\ \citenamefont {Yokozawa}}]{ref:Takeda2021}%
  \BibitemOpen
  \bibfield  {author} {\bibinfo {author} {\bibfnamefont {M.}~\bibnamefont {Takeda}}, \bibinfo {author} {\bibfnamefont {Y.}~\bibnamefont {Hiranuma}}, \bibinfo {author} {\bibfnamefont {N.}~\bibnamefont {Kanda}}, \bibinfo {author} {\bibfnamefont {K.}~\bibnamefont {Kotake}}, \bibinfo {author} {\bibfnamefont {T.}~\bibnamefont {Kuroda}}, \bibinfo {author} {\bibfnamefont {R.}~\bibnamefont {Negishi}}, \bibinfo {author} {\bibfnamefont {K.}~\bibnamefont {Oohara}}, \bibinfo {author} {\bibfnamefont {K.}~\bibnamefont {Sakai}}, \bibinfo {author} {\bibfnamefont {Y.}~\bibnamefont {Sakai}}, \bibinfo {author} {\bibfnamefont {T.}~\bibnamefont {Sawada}}, \bibinfo {author} {\bibfnamefont {H.}~\bibnamefont {Takahashi}}, \bibinfo {author} {\bibfnamefont {S.}~\bibnamefont {Tsuchida}}, \bibinfo {author} {\bibfnamefont {Y.}~\bibnamefont {Watanabe}},\ and\ \bibinfo {author} {\bibfnamefont {T.}~\bibnamefont {Yokozawa}},\ }\bibfield  {title} {\bibinfo {title} {Application of the {Hilbert-Huang} transform for analyzing
  standing-accretion-shock-instability induced gravitational waves in a core-collapse supernova},\ }\href {https://doi.org/10.1103/PhysRevD.104.084063} {\bibfield  {journal} {\bibinfo  {journal} {Phys. Rev. D}\ }\textbf {\bibinfo {volume} {104}},\ \bibinfo {pages} {084063} (\bibinfo {year} {2021})}\BibitemShut {NoStop}%
\bibitem [{\citenamefont {Powell}\ and\ \citenamefont {M\"{u}ller}(2019)}]{ref:Powell2019}%
  \BibitemOpen
  \bibfield  {author} {\bibinfo {author} {\bibfnamefont {J.}~\bibnamefont {Powell}}\ and\ \bibinfo {author} {\bibfnamefont {B.}~\bibnamefont {M\"{u}ller}},\ }\bibfield  {title} {\bibinfo {title} {Gravitational wave emission from {3D} explosion models of core-collapse supernovae with low and normal explosion energies},\ }\href {https://doi.org/10.1093/mnras/stz1304} {\bibfield  {journal} {\bibinfo  {journal} {Mon. Not. R. Astron. Soc.}\ }\textbf {\bibinfo {volume} {487}},\ \bibinfo {pages} {1178} (\bibinfo {year} {2019})}\BibitemShut {NoStop}%
\bibitem [{\citenamefont {Powell}\ and\ \citenamefont {M\"{u}ller}(2020)}]{ref:Powell2020}%
  \BibitemOpen
  \bibfield  {author} {\bibinfo {author} {\bibfnamefont {J.}~\bibnamefont {Powell}}\ and\ \bibinfo {author} {\bibfnamefont {B.}~\bibnamefont {M\"{u}ller}},\ }\bibfield  {title} {\bibinfo {title} {Three-dimensional core-collapse supernova simulations of massive and rotating progenitors},\ }\href {https://doi.org/10.1093/mnras/staa1048} {\bibfield  {journal} {\bibinfo  {journal} {Mon. Not. R. Astron. Soc.}\ }\textbf {\bibinfo {volume} {494}},\ \bibinfo {pages} {4665} (\bibinfo {year} {2020})}\BibitemShut {NoStop}%
\bibitem [{\citenamefont {M\"{u}ller}\ \emph {et~al.}(2010)\citenamefont {M\"{u}ller}, \citenamefont {Janka},\ and\ \citenamefont {Dimmelmeier}}]{ref:Muller2010}%
  \BibitemOpen
  \bibfield  {author} {\bibinfo {author} {\bibfnamefont {B.}~\bibnamefont {M\"{u}ller}}, \bibinfo {author} {\bibfnamefont {H.-T.}\ \bibnamefont {Janka}},\ and\ \bibinfo {author} {\bibfnamefont {H.}~\bibnamefont {Dimmelmeier}},\ }\bibfield  {title} {\bibinfo {title} {A new multi-dimensional general relativistic neutrino hydrodynamic code for core-collapse supernovae. {I}. {Method} and code tests in spherical symmetry},\ }\href {https://doi.org/10.1088/0067-0049/189/1/104} {\bibfield  {journal} {\bibinfo  {journal} {Astrophys. J. Suppl. Ser.}\ }\textbf {\bibinfo {volume} {189}},\ \bibinfo {pages} {104} (\bibinfo {year} {2010})}\BibitemShut {NoStop}%
\bibitem [{\citenamefont {Branchesi}\ \emph {et~al.}(2023)\citenamefont {Branchesi} \emph {et~al.}}]{ref:Branchesi2023}%
  \BibitemOpen
  \bibfield  {author} {\bibinfo {author} {\bibfnamefont {M.}~\bibnamefont {Branchesi}} \emph {et~al.},\ }\bibfield  {title} {\bibinfo {title} {{Science with the Einstein Telescope: A comparison of different designs}},\ }\href {https://doi.org/10.1088/1475-7516/2023/07/068} {\bibfield  {journal} {\bibinfo  {journal} {J. Cosmol. Astropart. Phys.}\ }\textbf {\bibinfo {volume} {2023}}\bibinfo  {number} { (07)},\ \bibinfo {pages} {068}}\BibitemShut {NoStop}%
\bibitem [{\citenamefont {Punturo}(2011)}]{ref:ET-D}%
  \BibitemOpen
\bibfield  {number} {  }\bibfield  {author} {\bibinfo {author} {\bibfnamefont {M.}~\bibnamefont {Punturo}},\ }\href@noop {} {\bibinfo {title} {{ET sensitivities page}}},\ \bibinfo {howpublished} {\href{https://www.et-gw.eu/index.php/etsensitivities}{https://www.et-gw.eu/index.php/etsensitivities}} (\bibinfo {year} {2011})\BibitemShut {NoStop}%
\bibitem [{\citenamefont {Hild}\ \emph {et~al.}(2011)\citenamefont {Hild} \emph {et~al.}}]{ref:Hild2011}%
  \BibitemOpen
  \bibfield  {author} {\bibinfo {author} {\bibfnamefont {S.}~\bibnamefont {Hild}} \emph {et~al.},\ }\bibfield  {title} {\bibinfo {title} {Sensitivity studies for third-generation gravitational wave observatories},\ }\href {https://doi.org/10.1088/0264-9381/28/9/094013} {\bibfield  {journal} {\bibinfo  {journal} {Classical Quantum Gravity}\ }\textbf {\bibinfo {volume} {28}},\ \bibinfo {pages} {094013} (\bibinfo {year} {2011})}\BibitemShut {NoStop}%
\bibitem [{\citenamefont {Moore}\ \emph {et~al.}(2014)\citenamefont {Moore}, \citenamefont {Cole},\ and\ \citenamefont {Berry}}]{ref:moore2014gravitational}%
  \BibitemOpen
  \bibfield  {author} {\bibinfo {author} {\bibfnamefont {C.~J.}\ \bibnamefont {Moore}}, \bibinfo {author} {\bibfnamefont {R.~H.}\ \bibnamefont {Cole}},\ and\ \bibinfo {author} {\bibfnamefont {C.~P.~L.}\ \bibnamefont {Berry}},\ }\bibfield  {title} {\bibinfo {title} {Gravitational-wave sensitivity curves},\ }\href@noop {} {\bibfield  {journal} {\bibinfo  {journal} {Classical and Quantum Gravity}\ }\textbf {\bibinfo {volume} {32}},\ \bibinfo {pages} {015014} (\bibinfo {year} {2014})}\BibitemShut {NoStop}%
\bibitem [{\citenamefont {Nitz}\ \emph {et~al.}(2024)\citenamefont {Nitz} \emph {et~al.}}]{ref:Pycbc}%
  \BibitemOpen
  \bibfield  {author} {\bibinfo {author} {\bibfnamefont {A.}~\bibnamefont {Nitz}} \emph {et~al.},\ }\href {https://doi.org/10.5281/ZENODO.596388} {\bibinfo {title} {gwastro/pycbc: v2.3.3 release of pycbc}} (\bibinfo {year} {2024})\BibitemShut {NoStop}%
\bibitem [{\citenamefont {Xu}\ \emph {et~al.}(2024)\citenamefont {Xu}, \citenamefont {Tiwari},\ and\ \citenamefont {Drago}}]{ref:Xu2024}%
  \BibitemOpen
  \bibfield  {author} {\bibinfo {author} {\bibfnamefont {Y.}~\bibnamefont {Xu}}, \bibinfo {author} {\bibfnamefont {S.}~\bibnamefont {Tiwari}},\ and\ \bibinfo {author} {\bibfnamefont {M.}~\bibnamefont {Drago}},\ }\bibfield  {title} {\bibinfo {title} {{PycWB}: {A} user-friendly, modular, and python-based framework for gravitational wave unmodelled search},\ }\href {https://doi.org/10.1016/j.softx.2024.101639} {\bibfield  {journal} {\bibinfo  {journal} {SoftwareX}\ }\textbf {\bibinfo {volume} {26}},\ \bibinfo {pages} {101639} (\bibinfo {year} {2024})}\BibitemShut {NoStop}%
\bibitem [{\citenamefont {Abbott}\ \emph {et~al.}(2020)\citenamefont {Abbott} \emph {et~al.}}]{ref:Abbott2020_ccsn}%
  \BibitemOpen
  \bibfield  {author} {\bibinfo {author} {\bibfnamefont {B.~P.}\ \bibnamefont {Abbott}} \emph {et~al.} (\bibinfo {collaboration} {LIGO Scientific Collaboration and Virgo Collaboration and ASAS-SN Collaboration and DLT40 Collaboration}),\ }\bibfield  {title} {\bibinfo {title} {{Optically targeted search for gravitational waves emitted by core-collapse supernovae during the first and second observing runs of Advanced LIGO and Advanced Virgo}},\ }\href {https://doi.org/10.1103/PhysRevD.101.084002} {\bibfield  {journal} {\bibinfo  {journal} {Phys. Rev. D}\ }\textbf {\bibinfo {volume} {101}},\ \bibinfo {pages} {084002} (\bibinfo {year} {2020})}\BibitemShut {NoStop}%
\bibitem [{\citenamefont {Necula}\ \emph {et~al.}(2012)\citenamefont {Necula}, \citenamefont {Klimenko},\ and\ \citenamefont {Mitselmakher}}]{ref:Necula2012}%
  \BibitemOpen
  \bibfield  {author} {\bibinfo {author} {\bibfnamefont {V.}~\bibnamefont {Necula}}, \bibinfo {author} {\bibfnamefont {S.}~\bibnamefont {Klimenko}},\ and\ \bibinfo {author} {\bibfnamefont {G.}~\bibnamefont {Mitselmakher}},\ }\bibfield  {title} {\bibinfo {title} {Transient analysis with fast wilson-daubechies time-frequency transform},\ }\href {https://doi.org/10.1088/1742-6596/363/1/012032} {\bibfield  {journal} {\bibinfo  {journal} {J. Phys. Conf. Ser.}\ }\textbf {\bibinfo {volume} {363}},\ \bibinfo {pages} {012032} (\bibinfo {year} {2012})}\BibitemShut {NoStop}%
\bibitem [{\citenamefont {Abbott}\ \emph {et~al.}(2016)\citenamefont {Abbott} \emph {et~al.}}]{ref:abbott2016}%
  \BibitemOpen
  \bibfield  {author} {\bibinfo {author} {\bibfnamefont {B.~P.}\ \bibnamefont {Abbott}} \emph {et~al.} (\bibinfo {collaboration} {LIGO Scientific Collaboration and Virgo Collaboration}),\ }\bibfield  {title} {\bibinfo {title} {Observing gravitational-wave transient {GW150914} with minimal assumptions},\ }\href {https://doi.org/10.1103/PhysRevD.93.122004} {\bibfield  {journal} {\bibinfo  {journal} {Phys. Rev. D}\ }\textbf {\bibinfo {volume} {93}},\ \bibinfo {pages} {122004} (\bibinfo {year} {2016})}\BibitemShut {NoStop}%
\bibitem [{\citenamefont {Wu}\ and\ \citenamefont {Huang}(2009)}]{ref:Wu2009}%
  \BibitemOpen
  \bibfield  {author} {\bibinfo {author} {\bibfnamefont {Z.}~\bibnamefont {Wu}}\ and\ \bibinfo {author} {\bibfnamefont {N.~E.}\ \bibnamefont {Huang}},\ }\bibfield  {title} {\bibinfo {title} {{Ensemble empirical mode decomposition: A noise-assisted data analysis method}},\ }\href {https://doi.org/10.1142/s1793536909000047} {\bibfield  {journal} {\bibinfo  {journal} {Adv. Adapt. Data Anal.}\ }\textbf {\bibinfo {volume} {01}},\ \bibinfo {pages} {1} (\bibinfo {year} {2009})}\BibitemShut {NoStop}%
\bibitem [{\citenamefont {Yeh}\ \emph {et~al.}(2010)\citenamefont {Yeh}, \citenamefont {Shieh},\ and\ \citenamefont {Huang}}]{ref:Yeh2010}%
  \BibitemOpen
  \bibfield  {author} {\bibinfo {author} {\bibfnamefont {J.-R.}\ \bibnamefont {Yeh}}, \bibinfo {author} {\bibfnamefont {J.-S.}\ \bibnamefont {Shieh}},\ and\ \bibinfo {author} {\bibfnamefont {N.~E.}\ \bibnamefont {Huang}},\ }\bibfield  {title} {\bibinfo {title} {{Complementary ensemble empirical mode decomposition: A novel noise enhanced data analysis method}},\ }\href {https://doi.org/10.1142/s1793536910000422} {\bibfield  {journal} {\bibinfo  {journal} {Adv. Adapt. Data Anal.}\ }\textbf {\bibinfo {volume} {02}},\ \bibinfo {pages} {135} (\bibinfo {year} {2010})}\BibitemShut {NoStop}%
\bibitem [{\citenamefont {Anderson}\ \emph {et~al.}(1999)\citenamefont {Anderson}, \citenamefont {Bai}, \citenamefont {Bischof}, \citenamefont {Blackford}, \citenamefont {Demmel}, \citenamefont {Dongarra}, \citenamefont {Du~Croz}, \citenamefont {Hammarling}, \citenamefont {Greenbaum}, \citenamefont {McKenney},\ and\ \citenamefont {Sorensen}}]{ref:LAPACK}%
  \BibitemOpen
  \bibfield  {author} {\bibinfo {author} {\bibfnamefont {E.}~\bibnamefont {Anderson}}, \bibinfo {author} {\bibfnamefont {Z.}~\bibnamefont {Bai}}, \bibinfo {author} {\bibfnamefont {C.}~\bibnamefont {Bischof}}, \bibinfo {author} {\bibfnamefont {L.~S.}\ \bibnamefont {Blackford}}, \bibinfo {author} {\bibfnamefont {J.}~\bibnamefont {Demmel}}, \bibinfo {author} {\bibfnamefont {J.~J.}\ \bibnamefont {Dongarra}}, \bibinfo {author} {\bibfnamefont {J.}~\bibnamefont {Du~Croz}}, \bibinfo {author} {\bibfnamefont {S.}~\bibnamefont {Hammarling}}, \bibinfo {author} {\bibfnamefont {A.}~\bibnamefont {Greenbaum}}, \bibinfo {author} {\bibfnamefont {A.}~\bibnamefont {McKenney}},\ and\ \bibinfo {author} {\bibfnamefont {D.}~\bibnamefont {Sorensen}},\ }\href@noop {} {\emph {\bibinfo {title} {LAPACK Users' guide (third ed.)}}}\ (\bibinfo  {publisher} {Society for Industrial and Applied Mathematics},\ \bibinfo {address} {USA},\ \bibinfo {year} {1999})\BibitemShut {NoStop}%
\bibitem [{\citenamefont {Bizouard}\ \emph {et~al.}(2021{\natexlab{b}})\citenamefont {Bizouard}, \citenamefont {Maturana-Russel}, \citenamefont {Torres-Forn{\'e}}, \citenamefont {Obergaulinger}, \citenamefont {Cerd{\'a}-Dur{\'a}n}, \citenamefont {Christensen}, \citenamefont {Font},\ and\ \citenamefont {Meyer}}]{ref:bizouard2021inference}%
  \BibitemOpen
  \bibfield  {author} {\bibinfo {author} {\bibfnamefont {M.-A.}\ \bibnamefont {Bizouard}}, \bibinfo {author} {\bibfnamefont {P.}~\bibnamefont {Maturana-Russel}}, \bibinfo {author} {\bibfnamefont {A.}~\bibnamefont {Torres-Forn{\'e}}}, \bibinfo {author} {\bibfnamefont {M.}~\bibnamefont {Obergaulinger}}, \bibinfo {author} {\bibfnamefont {P.}~\bibnamefont {Cerd{\'a}-Dur{\'a}n}}, \bibinfo {author} {\bibfnamefont {N.}~\bibnamefont {Christensen}}, \bibinfo {author} {\bibfnamefont {J.~A.}\ \bibnamefont {Font}},\ and\ \bibinfo {author} {\bibfnamefont {R.}~\bibnamefont {Meyer}},\ }\bibfield  {title} {\bibinfo {title} {Inference of protoneutron star properties from gravitational-wave data in core-collapse supernovae},\ }\href@noop {} {\bibfield  {journal} {\bibinfo  {journal} {Physical Review D}\ }\textbf {\bibinfo {volume} {103}},\ \bibinfo {pages} {063006} (\bibinfo {year} {2021}{\natexlab{b}})}\BibitemShut {NoStop}%
\bibitem [{\citenamefont {Sotani}\ \emph {et~al.}(2024)\citenamefont {Sotani}, \citenamefont {M{\"u}ller},\ and\ \citenamefont {Takiwaki}}]{ref:sotani2024universality}%
  \BibitemOpen
  \bibfield  {author} {\bibinfo {author} {\bibfnamefont {H.}~\bibnamefont {Sotani}}, \bibinfo {author} {\bibfnamefont {B.}~\bibnamefont {M{\"u}ller}},\ and\ \bibinfo {author} {\bibfnamefont {T.}~\bibnamefont {Takiwaki}},\ }\bibfield  {title} {\bibinfo {title} {Universality in supernova gravitational waves with protoneutron star properties},\ }\href@noop {} {\bibfield  {journal} {\bibinfo  {journal} {Physical Review D}\ }\textbf {\bibinfo {volume} {109}},\ \bibinfo {pages} {123021} (\bibinfo {year} {2024})}\BibitemShut {NoStop}%
\bibitem [{\citenamefont {Roberts}\ \emph {et~al.}(2012)\citenamefont {Roberts}, \citenamefont {Shen}, \citenamefont {Cirigliano}, \citenamefont {Pons}, \citenamefont {Reddy},\ and\ \citenamefont {Woosley}}]{ref:Roberts2012}%
  \BibitemOpen
  \bibfield  {author} {\bibinfo {author} {\bibfnamefont {L.~F.}\ \bibnamefont {Roberts}}, \bibinfo {author} {\bibfnamefont {G.}~\bibnamefont {Shen}}, \bibinfo {author} {\bibfnamefont {V.}~\bibnamefont {Cirigliano}}, \bibinfo {author} {\bibfnamefont {J.~A.}\ \bibnamefont {Pons}}, \bibinfo {author} {\bibfnamefont {S.}~\bibnamefont {Reddy}},\ and\ \bibinfo {author} {\bibfnamefont {S.~E.}\ \bibnamefont {Woosley}},\ }\bibfield  {title} {\bibinfo {title} {Protoneutron star cooling with convection: The effect of the symmetry energy},\ }\href {https://doi.org/10.1103/PhysRevLett.108.061103} {\bibfield  {journal} {\bibinfo  {journal} {Phys. Rev. Lett.}\ }\textbf {\bibinfo {volume} {108}},\ \bibinfo {pages} {061103} (\bibinfo {year} {2012})}\BibitemShut {NoStop}%
\bibitem [{\citenamefont {Kuroda}\ \emph {et~al.}(2016)\citenamefont {Kuroda}, \citenamefont {Kotake},\ and\ \citenamefont {Takiwaki}}]{ref:Kuroda2016}%
  \BibitemOpen
  \bibfield  {author} {\bibinfo {author} {\bibfnamefont {T.}~\bibnamefont {Kuroda}}, \bibinfo {author} {\bibfnamefont {K.}~\bibnamefont {Kotake}},\ and\ \bibinfo {author} {\bibfnamefont {T.}~\bibnamefont {Takiwaki}},\ }\bibfield  {title} {\bibinfo {title} {A new gravitational-wave signature from standing accretion shock instability in supernovae},\ }\href {https://doi.org/10.3847/2041-8205/829/1/L14} {\bibfield  {journal} {\bibinfo  {journal} {Astrophys. J. Lett.}\ }\textbf {\bibinfo {volume} {829}},\ \bibinfo {pages} {L14} (\bibinfo {year} {2016})}\BibitemShut {NoStop}%
\bibitem [{\citenamefont {Cornish}\ and\ \citenamefont {Littenberg}(2015)}]{ref:Cornish2015}%
  \BibitemOpen
  \bibfield  {author} {\bibinfo {author} {\bibfnamefont {N.~J.}\ \bibnamefont {Cornish}}\ and\ \bibinfo {author} {\bibfnamefont {T.~B.}\ \bibnamefont {Littenberg}},\ }\bibfield  {title} {\bibinfo {title} {{Bayeswave: Bayesian inference for gravitational wave bursts and instrument glitches}},\ }\href {https://doi.org/10.1088/0264-9381/32/13/135012} {\bibfield  {journal} {\bibinfo  {journal} {Classical Quantum Gravity}\ }\textbf {\bibinfo {volume} {32}},\ \bibinfo {pages} {135012} (\bibinfo {year} {2015})}\BibitemShut {NoStop}%
\bibitem [{\citenamefont {Cornish}\ \emph {et~al.}(2021)\citenamefont {Cornish}, \citenamefont {Littenberg}, \citenamefont {B\'ecsy}, \citenamefont {Chatziioannou}, \citenamefont {Clark}, \citenamefont {Ghonge},\ and\ \citenamefont {Millhouse}}]{ref:Cornish2021}%
  \BibitemOpen
  \bibfield  {author} {\bibinfo {author} {\bibfnamefont {N.~J.}\ \bibnamefont {Cornish}}, \bibinfo {author} {\bibfnamefont {T.~B.}\ \bibnamefont {Littenberg}}, \bibinfo {author} {\bibfnamefont {B.}~\bibnamefont {B\'ecsy}}, \bibinfo {author} {\bibfnamefont {K.}~\bibnamefont {Chatziioannou}}, \bibinfo {author} {\bibfnamefont {J.~A.}\ \bibnamefont {Clark}}, \bibinfo {author} {\bibfnamefont {S.}~\bibnamefont {Ghonge}},\ and\ \bibinfo {author} {\bibfnamefont {M.}~\bibnamefont {Millhouse}},\ }\bibfield  {title} {\bibinfo {title} {Bayeswave analysis pipeline in the era of gravitational wave observations},\ }\href {https://doi.org/10.1103/PhysRevD.103.044006} {\bibfield  {journal} {\bibinfo  {journal} {Phys. Rev. D}\ }\textbf {\bibinfo {volume} {103}},\ \bibinfo {pages} {044006} (\bibinfo {year} {2021})}\BibitemShut {NoStop}%
\end{thebibliography}%
\end{document}